\documentclass[journal,onecolumn]{IEEEtran}

\usepackage{cite}

\usepackage[cmex10]{amsmath}

\usepackage{amsmath,amssymb}

\def\RR{\rm \hbox{I\kern-.2em\hbox{R}}}
\def\NN{\rm \hbox{I\kern-.2em\hbox{N}}}
\def\ZZ{\rm {{\rm Z}\kern-.28em{\rm Z}}}
\def\CC{\rm \hbox{C\kern -.5em {\raise .32ex \hbox{$\scriptscriptstyle |$}}\kern -.22em{\raise .6ex \hbox{$\scriptscriptstyle |$}}\kern .4em}}

\def\<{\langle}
\def\>{\rangle}

\def\i{\infty}
\def\e{\varepsilon}

\def\Chi{\raise .3ex \hbox{\large $\chi$}} 
\def\[{\Bigl [}
\def\]{\Bigr ]}
\def\({\Bigl (}
\def\){\Bigr )}
\def\[{\Bigl [}
\def\]{\Bigr ]}
\def\({\Bigl (}
\def\){\Bigr )}

\def\b{\begin{equation}}
\def\e{\end{equation}}

\newcommand{\EE}[1]{{\mathbb E}\left[#1\right]}

\newcommand{\Var}[1]{{\mathbb V}\mathrm{ar}\left[#1\right]}

\newcommand{\Cov}[2]{{\mathbb C}\mathrm{ov}\left[#1,#2\right]}

\newtheorem{Nt}{Notation}
\newtheorem{Pp}{Proposition}
\newtheorem{Th}{Theorem}

\newtheorem{Lm}{Lemma}
\newtheorem{Cr}{Corollary}

\usepackage{graphics}
\usepackage{psfrag}
\usepackage{epsfig}

\usepackage{pifont}

\begin{document}

\title{Log-Normal continuous cascades:  aggregation properties and estimation.  \\
Application to financial time-series}

\author{E.~Bacry\thanks{E.Bacry~and A.~Kozhemyak are in the Centre de Math\'ematiques Appliqu\'ees, Ecole Polytechnique, 91128 Palaiseau, France.}, A. Kozhemyak and J.F.~Muzy\thanks{J.F.~Muzy is in the
Laboratoire Sciences Pour l'Environnement, CNRS, Universit\'e de Corse,
UMR 6134, 20250 Cort\'e, France.}}

\maketitle

\begin{abstract}
\noindent
Log-normal continuous random cascades form a class of multifractal processes that has already been successfully used in various fields. Several
statistical issues related to this model are studied.
We first make a quick but extensive review of their main properties and show that 
most of these properties can be analytically studied. 
We then develop an approximation theory of these processes 
in the limit of {\em small 
intermittency} $\lambda^2\ll 1$, i.e., when the degree of multifractality is small. This allows us to prove that the probability distributions associated
with these processes possess some very simple aggregation 
properties accross time scales. Such a control of the process properties at different time scales, allows us
to address the problem of parameter estimation. 
We show that one has to distinguish two different asymptotic regimes: the first one, referred to as  the ''low frequency regime'', corresponds to taking 
a sample whose overall size increases whereas the second one, referred to as  the ''high frequency regime'', corresponds to sampling the process at an increasing sampling rate. 
We show that, the first regime leads to convergent estimators whereas, 
in the high frequency regime, the situation is much more intricate : only 
the intermittency coefficient $\lambda^2$ can be estimated using a consistent estimator. 
However, we show that, in practical situations, one can detect the nature of the asymptotic regime (low frequency versus high frequency) and consequently decide whether the estimations of the other parameters are reliable or not.
We finally illustrate how both our results on parameter estimation and on aggregation properties, allow one 
to successfully use these models for modelization and 
prediction of financial time series.
\end{abstract}


\section{Introduction}
\label{1}
\PARstart
Data displaying multi-scaling behavior are observed in various fields of applied and fundamental sciences:
the velocity field of fully developed turbulent flows \cite{Fri95}, financial time-series \cite{BouPot03, BSM}, the
telecommunication traffic load in high speed networks \cite{aPa98}, medical time-series
\cite{bWe90,aWe94}, geological shapes \cite{bWi95} are only few of numerous
examples. The paradigm of multifractal processes are multiplicative cascades originally 
introduced by the russian school \cite{Kol62} for modelling the energy 
cascade in fully developed turbulence and further studied by Mandelbrot \cite{Man74a,Man74b}. 
Very recently, continuous versions of these processes have been
defined: they share exact multifractal scaling with discrete cascades
but they display continuous scaling and possess stationary
increments \cite{MuzDelBac00,BarMan02,MuzBac02,BacMuz03}.  
Despite the huge number of  mathematical studies devoted to discrete (e.g., \cite{KahPey76,Gui87,Mol96,Mol97,Liu02})
or continuous random cascades \cite{BarMan02,BacMuz03,ChaRieAbr03,Bar04,BarMan04a},
only very few works considered standard statistical problems associated with these processes 
(see however 
\cite{OssWay00, ResSamGilWil03, Lux03, Lux04, CalFis04}).
Our goal in this paper is to adress several statistical issues related 
to multifractal processes. 

The self-similarity of a process $X(t)$ \footnote{We consider exclusively processes $X(t)$ with stationary
increments and such that $X(0) = 0$. In that respect, $\forall t'$, $X(t'+t)-X(t')$, has the same distribution than $X(t)$.} can be characterized by 
the power-law behavior of the $q$-order moments of its increments as functions of
the scale 
\begin{equation}
\forall q \in \RR,~~\EE{|X(t)|^{q}} \simeq  C_{q} t^{\zeta_X(q)},~\forall t\le T,
\end{equation}
where $T$ is referred to as the ``integral scale'', it actually corresponds to a decorrelation scale.
In the case the so-obtained ``scaling
exponents'' $\zeta_X(q)$ are not depending linearly on $q$ but
is a  concave function of $q$, the process is said to be a multifractal process. Since scaling of 
moments of different orders do not behave homogeously as the time scale is changed, 
the probability distribution of the increments of a multifractal process strongly depends on the scale of the increments.For random cascade models, 
one can show that the scaling exponent, as a function of $q$, corresponds 
to the cumulant generating function of a log-infinitely divisible law $W$,
\begin{equation}
\label{eq1}
\zeta_X(q) = \ln \EE{W^q}.
\end{equation} 
Given the infinitely divisible law $W$, 
the continuous cascade process is entirely defined as soon as its variance $\sigma^2$ (i.e., a simple multiplicative factor) is fixed, as well as its integral scale $T$ (i.e., the ''size'' of the cascade, or the decorrelation scale).
In this paper, we will exclusively 
focus on continuous cascades with log-normal scaling exponents. 

Among the whole 
class of log infinitely divisible cascade models, log normal cascades have the advantage 
of being fully determined by a single parameter $\lambda^2$, corresponding basically to the variance of $\ln W$. Thus, as shown by Eq. \eqref{eq1}, this parameter $\lambda^2$ 
rules the non linearity of the scaling exponents (i.e., the ``degree'' of multifractality of the process). It is referred to as the
the intermittency coefficient. 
Moreover, log-normal continuous cascades have a very simple alternative construction and, as we will see in the next section  of this paper, most of their  properties 
can be expressed under closed formulae. 
 Though, they correspond to a particular log-infinitely divisible law, they are rich enough to 
raise challenging statistical questions such as estimator convergence in the context of long-memory correlation. 
Our approach relies upon an approximation that allows us to precisely
control the aggregation properties of the process. In fact, we show
that in the small intermittency limit ($\lambda^2 \ll 1$), a log-normal continuous cascade has
increments that are ``close'' to be log-normal, at each scale, in a sense that will 
be precisely defined in the sequel.  
This approximation framework allows us to develop a method to estimate
the process parameters. In that context, we are lead to introduce 
two distinct situations for asymptotic regimes: the first one, referred to as the ''low frequency regime'',  corresponds to 
the classical notion of infinite observation scale at a fixed sampling rate while the second, referred to as the ''high frequency regime'',   corresponds to sampling the process over a fixed observation scale at an increasing sampling rate. 
More precisely, if  $\tau$ is the sampling rate and  $L$ is the observation scale, the observed samples corresponds to the values
\begin{equation}
\{X(n\tau)\}_{n\in [0,N[},~~\mbox{where~} N = \frac L \tau.
\end{equation}
Both asymptotic regimes corresponds to $N \rightarrow +\infty$, however, whereas the low-frequency regime corresponds to $\tau$ fixed and $L\rightarrow +\infty$, the high-frequency regime corresponds to $L$ fixed and $\tau \rightarrow 0$. From an experimental point of view, the first regime corresponds to the case where $\frac L T \gg 1$ and $\frac T \tau \simeq 1$ whereas the second regime corresponds to the case where
$\frac L T \simeq 1$ and $\frac T \tau \gg 1$.
We show that the properties of the parameter estimators are fundamentally different depending on the nature of the asymptotic regime 
(and allow, by the way, to test what the effective nature of the regime is). In the last section, we apply all these results on parameter estimation for the calibration of a multifractal model
to account for volatility dynamics in financial time series. Moreover, making extensive use of the aggregation properties, we show that the so-obtained model provides highly performant methods to forecast risk.

The paper is organized as follows: in section \ref{2} we recall the main
definition of log-normal cascades at the heart of this study and we
state its main properties.  We study both the case of the Multifractal Random Measure (MRM), a non-decreasing process, and the case of the Multifractal Random Walk (MRW), a Brownian motion subordinated by the MRM. 
The aggregation properties
of the model are discussed extensively in section \ref{sil} where we develop our 
small intermittency approximation theory. In this section, we first introduce a Gaussian process : the ``renormalized magnitude'' $\Omega(t)$ that will be involved in all the following approximations.
We show that, in the case $\lambda^2 \ll 1$, in some sense to be defined, the variations of the MRM or of the MRW are closely related to those of   $\Omega(t)$. Whereas subsection \ref{ssec:law} states a convergence theorem of the logarithm of the MRM towards $\Omega$ in the limit $\lambda^2 \rightarrow 0$,  the other subsections establish different MRM/MRW moment approximation theorems (when $\lambda^2 \ll 1$) as functions of  $\Omega(t)$ moments.
In section \ref{sec:est}, we show
how these approximations can be used to calibrate the model. The estimation
issues are discussed within both low-frequency and high-frequency
asymptotic regimes. 
We first show, in Section \ref{sec:GMM}, that, in the low-frequency regime, the 
``Generalized Moments Method'' (GMM) leads to convergent estimators whereas, as shown in Section \ref{sec:estgrT}, in the high-frequency regime, the situation is more intricate.
Indeed, in this regime, the integral scale $T$ is shown to be a ``fake'' parameter and $\sigma^2$ cannot be estimated. However, in experimental situation, 
the order of the GMM estimation of the fake parameter $T$ is proved to give some hints about the nature of the asymptotic regime and consequently about the reliability of the estimation of $T$ and $\sigma^2$. In Section \ref{sec:estlambda2}, we exhibit a GMM type estimator of $\lambda^2$ that is proved to be consistent. Numerical experiments illustrate all the estimation results.
In Section \ref{apps} we apply our results concerning parameter estimation tp the calibration of a model for
asset price fluctuations in financial markets. We then show how the aggregation properties can be successfully used to perform conditional risk 
forecasting. Conclusions and prospects are provided in section \ref{cp}
while Appendices contain additional technical results.

\section{Definitions and main properties of log-normal continuous cascade models}
\label{2}

In this section we will first 
focus on the definition of Multifractal Random Measures (MRM) originially introduced in \cite{BacMuz03}.
We will denote $M[t_0,t_1]$ the measure of the interval $[t_0,t_1]$ and $M(t)$ the non decreasing process $M(t) = M[0,t]$.
We propose two different approaches 
to define log-normal MRM. The first one 
relies upon the construction of some temporal Gaussian process
which covariance mimics the observed ``ultrametric'' covariance
of discrete Mandelbrot cascades while the second one involves
random measures in a 2D half-plane. The direct construction has 
the advantage of being simpler and easy to implement while the 
second construction can be easliy extended to other infinitely divisible
laws than the Gaussian law. 
Multifractal Random Walks (MRW) can be easily obtained from MRM by
compounding a self-similar stochastic process with $M(t)$.

\subsection{Direct definition}
\label{dd}
Let the measure 
$M_{l,T}(dt)$ be defined by
\b
\label{defomega}
  M_{l,T}(dt) = e^{2 \omega_{l,T}(t)} dt \; .
\e
in the sense that for all Lebesgue measurable set $I$ one has, $M_{l,T}(I) = \int_{I}e^{2 \omega_{l,T}(t)} dt$.
The process $\omega_{l,T}(t)$ is Gaussian and stationary and is defined by its mean and 
covariance function :
\b
\label{def:mean}
\EE{\omega_{l,T}(t)} = -\lambda^2 \left(\ln\left(\frac{T}{l} \right)+1 \right)
\e 
and 
\begin{equation}
\label{def:magcov}
\rho_{l,T}(\tau)={\mathbb C}{\mathrm ov}\big[\omega_{l,T}(t),\omega_{l,T}(t+\tau)\big]=
\begin{cases}
\lambda^2 \left(\ln\left(\frac{T}{l}\right)+1-\frac{\tau}{l}\right), \; \; \text{if $0\leq\tau< l$},
\\
\lambda^2 \ln\left(\frac{T}{\tau}\right), \; \; \text{if $l\leq\tau<T$},
\\
0, \; \; \text{if $T\leq\tau<+\infty$},
\end{cases}
\end{equation}
where the parameters $T$ is the integral scale and $\lambda^2$ is the intermittency coefficient.
Note that the fact that expression \eqref{def:magcov} represents of definite positive function is
proven in the next section.
Using Kahane Chaos theory \cite{Kah87}, on can prove that the weak limit
\begin{equation}
M_T(dt)=\lim_{l \to 0^{+}}M_{l,T}(dt),
\end{equation}
exists and is non trivial as long as $\lambda^2 < 1/2$.

Let us remark that the process $\{\omega_{l,T}(t)\}_t$ can be represented 
as a stochastic integral of a Kernel against the Wiener White noise $dB(t)$, like 
Mandelbrot-Van Ness fractional Brownian motion representation:
\begin{equation}
\omega_{l,T}(t)=-\EE{\omega_{l,T}(t)}+\int_{-\infty}^{t}K_{l,T}(t-u)dB(u),
\end{equation}
where the kernel $K_{l,T}$ (that can be chosen to be causal)
satisfies the convolution equation:
\begin{equation}
K_{l,T}\ast K_{l,T}(t)={\mathbb C}{\mathrm ov}\big[\omega_l(0),\omega_l(t)\big].
\end{equation}
A simple Fourier transform of this equation together with expression \eqref{def:magcov} allows one to show that the process $\omega_{l,T}$ can be 
seen as a kind of fractional Brownian motion in the marginal 
limit $H \rightarrow 0$. Indeed, the kernel $K_{l,T}(t)$ in previous equation
behaves, in the range $l\ll t\ll T$, like:
\begin{equation}
K_l(t)\sim\frac{K_0}{\sqrt{t}}.
\end{equation}
For this reason, as emphasized in ref. \cite{MuzBac02}, the MRM measure can be loosely defined as the exponential of an $1/f$ noise.

\subsection{Alternative definition : continuous cascades}
\label{gc}
The previous construction is hard to extend 
to other laws than the Gaussian law. A more general construction
that allows one to build continuous cascades with
arbitrary log-infinitely divisible statistics has been
proposed by Bacry and Muzy \cite{BacMuz03}. It amounts in building the process $\omega_{l,T}(t)$ from a 2d representation.

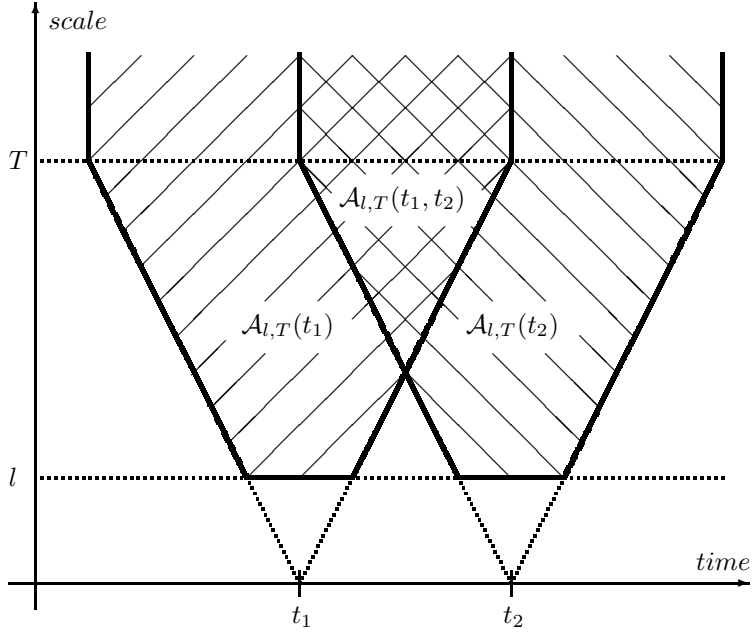
\begin{figure}
\begin{center}
\begin{picture}(300,260)
\put(10,30){\vector(1,0){280}}
\put(20,20){\vector(0,1){230}}
\put(270,35){$time$}
\put(25,240){$scale$}

\linethickness{0.8pt}
\put(120,25){\line(0,1){10}}
\put(200,25){\line(0,1){10}}
\put(117,15){$t_1$}
\put(197,15){$t_2$}
\put(10,67){$l$}
\put(10,187){$T$}
\put(98,125){${\mathcal A}_{l,T}(t_1)$}
\put(183,125){${\mathcal A}_{l,T}(t_2)$}
\put(135,173){${\mathcal A}_{l,T}(t_1,t_2)$}

\put(160,110){\line(1,-1){40}}
\put(140,150){\line(1,-1){80}}

\put(120,190){\line(1,-1){12}}
\put(145,165){\line(1,-1){35}}
\put(193,117){\line(1,-1){34}}

\put(120,210){\line(1,-1){25}}
\put(165,165){\line(1,-1){28}}
\put(213,117){\line(1,-1){21}}

\put(120,230){\line(1,-1){45}}
\put(185,165){\line(1,-1){28}}
\put(226,124){\line(1,-1){15}}

\put(140,230){\line(1,-1){107}}
\put(160,230){\line(1,-1){94}}
\put(180,230){\line(1,-1){80}}
\put(200,230){\line(1,-1){67}}
\put(220,230){\line(1,-1){54}}
\put(240,230){\line(1,-1){40}}
\put(260,230){\line(1,-1){20}}

\put(160,110){\line(-1,-1){40}}
\put(180,150){\line(-1,-1){80}}

\put(200,190){\line(-1,-1){11}}
\put(175,165){\line(-1,-1){35}}
\put(127,117){\line(-1,-1){34}}

\put(200,210){\line(-1,-1){25}}
\put(155,165){\line(-1,-1){28}}
\put(107,117){\line(-1,-1){21}}

\put(200,230){\line(-1,-1){45}}
\put(137,167){\line(-1,-1){30}}
\put(92,122){\line(-1,-1){13}}

\put(180,230){\line(-1,-1){107}}
\put(160,230){\line(-1,-1){94}}
\put(140,230){\line(-1,-1){80}}
\put(120,230){\line(-1,-1){67}}
\put(100,230){\line(-1,-1){54}}
\put(80,230){\line(-1,-1){40}}
\put(60,230){\line(-1,-1){20}}

\linethickness{1.5pt}
\qbezier(40,230)(40,210)(40,190)
\qbezier(40,190)(70,130)(100,70)
\qbezier(100,70)(120,70)(140,70)
\qbezier(140,70)(170,130)(200,190)
\qbezier(200,190)(200,210)(200,230)

\qbezier(120,230)(120,210)(120,190)
\qbezier(120,190)(150,130)(180,70)
\qbezier(180,70)(200,70)(220,70)
\qbezier(220,70)(250,130)(280,190)
\qbezier(280,190)(280,210)(280,230)

\linethickness{1pt}
\qbezier[100](20,70)(150,70)(280,70)
\qbezier[100](20,190)(150,190)(280,190)

\qbezier[20](100,70)(110,50)(120,30)
\qbezier[20](120,30)(130,50)(140,70)

\qbezier[20](180,70)(190,50)(200,30)
\qbezier[20](200,30)(210,50)(220,70)

\end{picture}
\end{center}
\caption[Cone-like domains ${\mathcal A}_{l,T}(t_1)$, ${\mathcal A}_{l,T}(t_2)$ 
et ${\mathcal A}_{l,T}(t_1,t_2)$ used in the definition of the 
log-infinitely divisible continuous cascades]
{\small Cone-like domains ${\mathcal A}_{l,T}(t_1)$, ${\mathcal A}_{l,T}(t_2)$ 
et ${\mathcal A}_{l,T}(t_1,t_2)$ used in the definition of continuous
cascades (see definition \ref{eq:conMRM}). The parameter $T$ is the
integral scale and the parameter $l$ is the small scale cut-off. The limit 
MRM is obtained in the limit $l \rightarrow 0$.}
\label{fig1}
\end{figure}

We distribute a non centered gaussian white noise ${\mathcal P}$ of variance $\lambda^2$ on the  half plane $\left\{(t,l);t\in{\mathbb R},l\in{\mathbb R}^{+*}\right\}$ using the density measure $\mu(dt,dl)=l^{-2}dtdl$. Consequently, for any measurable set ${\mathcal A}$ of the half-plane, ${\mathcal P}({\mathcal A})$ is a gaussian random variable whose Laplace transform is of the form
\begin{equation}
\label{fcp}
{\mathbb E}\left[e^{q{\mathcal P}({\mathcal A})}\right]=e^{\psi(q)\mu({\mathcal A})}.
\end{equation}
If we choose the mean of the white noise ${\mathcal P}$ such that for any $ {\mathcal A} \in  {\mathcal S}^+$, one has $ {\mathbb E}[{\mathcal P}(A)] = -\Var{{\mathcal P}({\mathcal A})} = -2\lambda^2\mu({\mathcal A})$, then
\begin{equation}
\psi(q) = 2\lambda^2 q^2 -2\lambda^2q.
\end{equation}
Then if we define, for all $l$ and $T$ such that $0<l<T$, the cone-like domain:
${\mathcal A}_{l,T}(t)$ as:
\begin{equation}
\label{eq:conMRM}
{\mathcal A}_{l,T}(t)=\Big\{(t',l');l\leq l',|t'-t|\leq \frac{1}{2}\min(l',T)\Big\},
\end{equation}
the gaussian process $\omega_{l,T}(t)$ defined by Eqs (\ref{def:mean}) and (\ref{def:magcov}) has the following representation
(see fig. \ref{fig1})
\begin{equation}
\label{def:mag}
\omega_{l,T}(t)=\frac{1}{2}{\mathcal P}({\mathcal A}_{l,T}(t)),
\end{equation}

\subsection{Properties}

Many properties of the MRM measure involves the following concave function
\begin{equation}
\label{def:zetam}
  \zeta_M(q) = q - \psi(q) = (1+2\lambda^2)q-2\lambda^2q^2 \; ,
\end{equation}
which satisfies $\zeta_{M}(0) = 0$ and $ \zeta_{M}(1) = 1$.
The non degeneracy condition $\lambda^2 < 1/2$, can be rewritten in terms of $\zeta_{M}(q)$ : 
\begin{Pp}[Non degeneracy of $M$ \cite{BacMuz03}]
\begin{equation}
\lambda^2 < \frac 1 2 \Longleftrightarrow \zeta_{M}'(1) > 0 \Longrightarrow M {\mbox {~is non degenerated and~}} \forall t,~{\mathbb E}\left[M[0,t]\right] = t
\end{equation}
\end{Pp}
The limit measure $M$ possesses exact scale invariance properties that are directly resulting from the invariance
of the function $\rho_{l,T}(\tau)$ as respect to time dilation. Indeed,
the covariance function $\rho_{l,T}(\tau)$ in Eq. (\ref{def:magcov}) satisfies the following invariance properties
\begin{itemize}
\item[(i)] $\rho_{sl,sT}(s\tau) =  \rho_{l,T}(\tau),~\forall s > 0$,
\item[(ii)] $\rho_{l,T}(s\tau) =  \rho_{l,sT}(\tau)-\ln \lambda ,~\forall \tau\le sT, s \in [0,1]$,
\item[(iii)] $\rho_{sl,T}(s\tau) =  \rho_{l,T}(\tau)-\ln \lambda ,~\forall \tau\le T, s \in [0,1]$,
\end{itemize} 
It follows that the MRM measure $M$ satisfies 3 scale invariance properties 
\begin{Pp}[Scale invariance properties \cite{BacMuz03}] 
\label{prop:scaleinv}
~\\
\begin{itemize}
\item[(i)] Global scale-invariance
\begin{equation}
\label{eq:gscaleinv}
\{M_{sT}[0,s t]\}_t\overset{{\mathcal L}}{=}s\{M_T[0,t]\}_t,~\text{$\forall s \in \RR^+$} 
\end{equation}
\item[(ii)] Integral scale invariance property
\begin{equation}
\label{eq:iscaleinv}
\{M_T[0,t]\}_{0\le t\le sT}\overset{{\mathcal L}}{=}W_{s}\{M_{s T}[0,t]\}_{0\le t\le sT},~\text{$\forall s \in [0,1]$} 
\end{equation} 
\item[(iii)] Stochastic scale-invariance property
\begin{equation}
\label{eq:propas}
\{M_T[0,s t]\}_{0\le t\le T}\overset{{\mathcal L}}{=}W_{s}\{M_T[0,t]\}_{0\le t\le T},~\text{$\forall s \in [0,1]$}
\end{equation} 
\end{itemize} 
with, in the last two equations,
$W_{s}= s e^{\Omega_{s}}$, where $\Omega_{s}$ is a gaussian variable independant of $M$ and defined by
$\EE{\Omega_{s}} = -\Var{\Omega_{s}}/2 = 2\lambda^2 \ln s$
\end{Pp}
From  (iii) one can easily deduce that q-order moments verify an exact scale invariance property. Actually, one can show that 
\begin{Pp}[Finiteness of positive order moments \cite{BacMuz03}]
\label{thm:momM}
Let $\lambda^2<1/2$.
Let $q>0$. If $\zeta_M(q)>1$ then $\EE{M_T[0,t]^q}< +\infty$. Conversely, let
$q>1$, if $M_T \neq 0$ then 
$\EE{M_T[0,t]^q}<+\infty \Rightarrow \zeta_M(q) \geq 1$.
\end{Pp}
Moreover, a straightforward computation shows that
\begin{equation}
\zeta_M(q) = \frac {\log \EE{ {W_s} }} {\ln s},
\end{equation}
where $W_s$ is defined in Proposition \ref{prop:scaleinv}(iii)
One thus gets
\begin{Pp}[Exact scale invariance of $q$-order moments \cite{BacMuz03}]
Let $\lambda^2<1/2$. 
\begin{equation}
\label{eq:momMRW}
\forall q \in \RR,~~\EE{M_T[0,t]^{q}} =  K_{q} t^{\zeta_M(q)},~\forall t\le T,
\end{equation}
where the prefactor $K_{q}$ has an analytic formula in the case $q = n \in \NN^*$ : 
\begin{equation}
\label{eq:K}
K_{n}= \sigma^{2n}T^{2n(n-1)\lambda^2}
\prod_{k=0}^{n-1}\frac{\Gamma(1-2(k+1)\lambda^2)\Gamma(1-2k\lambda^2)^2}
{\Gamma(2-2(p+k-1)\lambda^2)\Gamma(1-2\lambda^2)}.
\end{equation}
\end{Pp}
Moreover one can prove that all the negative order moments exist
\begin{Pp}[Finiteness of negative order moments (condition (C4) in \cite{Bar04})]
\label{thm:mmomM}
Let $\lambda^2<1/2$. Then, $\forall q< 0$ we have $\EE{M_T[0,t]^q}< +\infty$. 
\end{Pp}

\subsection{The Multifractal Random Walk model}

As said previously, a large class of multifractal stochastic processes
can be associated with a given MRM. The simplest way is probably  
the approach initiated by Mandelbrot and Taylor \cite{ManTay67} that consists in
compounding a self-similar stochastic process 
which increments are stationnary with the
non decreasing function $M_T[0,t]$, where $M_T$ is a MRM as build in previous section. 
Another approach, inspired from econometrics, is to 
consider the measure $M_T(dt)$ as a stochastic variance associated with
a Brownian motion \cite{BacDelMuz01,MuzBac02}. 
In this paper, for the sake of simplicity and concision, 
we will exclusively consider processes with stationnary and 
uncorrelated increment constructed from the standard Brownian motion.
We define a  Multifractal
Random Walk (MRW) as follows:
Let $M_T$ be a (log-normal) MRM and consider $B(t)$ a Brownian motion\footnote{Note that 
a simple way to introduce long-range correlations in the MRW model would be to replace
the Brownian motion $B(t)$ by a fractional Brownian motion $B_H(t)$.} independent 
of $M_T$. The MRW $X_T(t)$ is simply defined as, for all $t \geq 0$:
\begin{equation}
\label{def:mrw}
   X_T(t) = B \Big( M_T[0,t] \Big)
\end{equation}

An alternative construction is obtained by considering the
stochastic integral of the measure $M_{l,T}$ as respect to Wiener measure $dB(u)$ (independent of $M_{l,T}$) and 
then take the (weak) limit $l \rightarrow 0$: 
\begin{equation}
X_T(t)=\lim_{l\to 0^{+}} \int\limits_{0}^{t}e^{\omega_{l,T}(u)}dB(u).
\end{equation}
Let us note that the equivalence between these two definitions is proven 
in \cite{BacMuz03}

The properties of the MRW 
directly result from those of the MRM and the self-similarity
of Brownian motion. 
\begin{Pp}[Scale invariance properties \cite{BacMuz03}] 
\label{prop:scaleinvmrw}
~\\
\begin{itemize}
\item[(i)] Global scale-invariance
\begin{equation}
\{X_{sT}(s t)\}_t\overset{{\mathcal L}}{=}s\{X_T(t)\}_t,~\text{$\forall s \in \RR^+$} 
\end{equation}
\item[(ii)] Integral scale invariance property
\begin{equation}
\label{eq:iscaleinvmrw}
\{X_T(t)\}_{0\le t\le sT}\overset{{\mathcal L}}{=}W_{s}\{X_{s T}(t)\}_{0\le t\le sT},~\text{$\forall s \in [0,1]$} 
\end{equation} 
\item[(iii)] Stochastic scale-invariance property
\begin{equation}
\label{eq:propasmrw}
\{X_T(st)\}_{0\le t\le T}\overset{{\mathcal L}}{=}W_{s}\{X_T(t)\}_{0\le t\le T},~\text{$\forall s \in [0,1]$} 
\end{equation} 
\end{itemize} 
with, in the last two equations,
$W_{s}= s^{1/2} e^{\Omega_{s}/2}$, where $\Omega_{s}$ is a gaussian variable independant of $M$ and defined by
$\EE{\Omega_{s}} = -\Var{\Omega_{s}}/2 = 2\lambda^2 \ln s$
\end{Pp}
\begin{Pp}[Finiteness of positive order moments \cite{BacMuz03}]
\label{thm:momX}
Let $\lambda^2<1/2$. Let
\begin{equation} 
\label{def:zetax}
\zeta_{X}(q)=\zeta_{M}(\frac{q}{2}) = \frac{q}{2}(1+2\lambda^2)-\frac{\lambda^2}{2}q^2 \; .
\end{equation}
Let $q>0$. If $\zeta_X(q)>1$ then $\EE{X_T(t)^q}< +\infty$. Conversely, let
$q>1$, if 
$\EE{X_T(t)^q}<+\infty \Rightarrow \zeta_X(q) \geq 1$.
\end{Pp}
A straightforward computation shows that
\begin{equation}
\zeta_X(q) = \frac {\log \EE{ {W_s} }} {\ln s},
\end{equation}
where $W_s$ is defined in Proposition \ref{prop:scaleinvmrw}(iii).
One thus gets
\begin{Pp}[Exact scale invariance of $q$-order moments \cite{BacMuz03}]
\label{prop:escaleinv}
Let $\lambda^2<1/2$. 
\begin{equation}
\label{eq:momX}
\forall q \in \RR,~~\EE{X_T(t)^{q}} =  \tilde K_{q} t^{\zeta_X(q)},~\forall t\le T,
\end{equation}
where the prefactor $\tilde K_{q}$ has an analytic formula in the case $q = n \in \NN$ : 
$\tilde K_{n} = (2n-1)!! \; K_{n}$, where $K_{n}$ is given by (\ref{eq:K}).
\end{Pp}
\begin{Pp}[Finiteness of negative order moments (condition (C4) in \cite{Bar04})]
\label{thm:mmomX}
Let $\lambda^2<1/2$. Then, $\forall q< 0$ we have $\EE{X_T(t)^q}< +\infty$. 
\end{Pp}

\subsection{Discrete time representation of a MRW - Monte Carlo simulation}
\label{montecarlo}
\begin{Nt}
For the sake of simplicity, in the following, if $Y(t)$ is a sochastic process, we will use the notation
\begin{equation}
\delta_\tau Y(t) = Y(t)-Y(t-\tau).
\end{equation}
Moreover we recall that if $M(dt)$ is a measure, $M(t)$ refers to the non decreasing process
\begin{equation}
M(t) = M[0,t],
\end{equation}
and, consequently
\begin{equation}
\delta_\tau M(t) = M(t)-M(t-\tau) = M[t-\tau,t].
\end{equation}
\end{Nt}
Let fix $\tau > 0$. We want to simulate the discrete time process $\{X_T(n\tau)\}_n$.
Approximated Monte Carlo simulation of this discrete time process can be obtained using 
Eqs \eqref{defomega}, \eqref{def:mean} and \eqref{def:magcov}. One first fixes $l$ small enough ($l=\frac \tau {128}$ will be sufficient for the purpose of this paper) such that $\frac \tau l$ is an integer.  The Gaussian stationary discrete time process $\{\omega_{l,T}(n\tau)\}_n$ can be simulated using the analytical formulae of its mean 
\eqref{def:mean} and of its autocovariance  \eqref{def:magcov}. 

Thus one can easily simulate  the measure $\tilde M_{l,T}(dt)$  that is uniform on each interval of the form $[kl,(k+1)l]$ with the density $e^{2\omega_{l,T}(kl)}$.
For $n\ge 0$, one has
\begin{equation}
\label{mtilde}
 \tilde M_{l,T}(nl) = \tilde M_{l,T}[0,nl] = \sum_{k=0}^{n-1} e^{2\omega_{l,T}(kl)} l.
\end{equation}
From these simulations, one can easily simulate the   process $\{\tilde X_{l,T}(t)\}_{t\ge 0}$ which is linear on each interval of the form $[kl,(k+1)l]$, and which satisfies
\begin{equation}
\label{xtilde}
\tilde X_{l,T}(nl) = \sum_{k=1}^{n} \epsilon[k]\sqrt{l} e^{\omega_{l,T}(k)},
\end{equation}
where $\epsilon[k]$ is a gaussian white noise which is independant of $\tilde M$.
The convergence of the linear-wise process $M_{l,T}(t)$ towards $M(t)$ when $l\rightarrow 0$, and consequently the convergence of 
the linear-wise process  $X_{l,T}(t)$ towards $X_T(t)$, are proved in \cite{BacMuz03}. 
Thus, simulations of the discrete-time process $\{X_{l,T}(n\tau)\}_n$ can be seen as a good approximations of simulations of the discrete-time process
$\{X_T(n\tau)\}_n$.

\section{Aggregation properties}
\label{sil}

\subsection{Introduction}
One of the nice features of standard Brownian motion 
is its stability as respect to time aggregation: 
At each scale, the increment probability distributions 
remain Gaussian. 
Propositions \ref{prop:scaleinv}(iii) and \ref{prop:scaleinvmrw}(iii) state that both the log-normal MRM and MRW processes
have stochastic scale-invariance property. This means that, in some sense, they possess stable 
properties when changing the time scale. 
However, this property is of poor practical interest because it does 
not provide the probability law at a given time scale $\tau$ 
but simply indicates how this law changes as $\tau$ 
varies.  In the log-normal continuous cascade models, the multifractality, i.e.,  
the non-linearity of the moment scaling exponent $\zeta_M(q)$,
is fully characterized by the intermittency coefficient $\lambda^2$.
Empirically this exponent is often found to be close to zero:
For instance, the commonly reported  value of $\lambda^2$ for
energy dissipation field Eulerian Turbulence and for
the volatility fluctuations associated with financial asset returns are
respectively $\lambda^2 \simeq 0.2$ and $\lambda^2 \simeq 0.02$.
It is therefore natural to study the properties of the log-normal MRM
measure in the limit $\lambda^2 \ll 1$.
In this section, we will see how, in this regime, the law of the MRM can be well approximated by the law of an explicit log normal process based on the so-called 
(normal) {\em renormalized magnitude} process $\Omega(t)$. 

In this section, we show that, in this regime, the variations of the MRM or of the MRW are closely related to those of   an explicit log normal process based on the so-called (normal) {\em renormalized magnitude} process $\Omega(t)$. 
 Whereas subsection \ref{ssec:law} states a convergence theorem of the logarithm of the MRM towards $\Omega$ in the limit $\lambda^2 \rightarrow 0$,  the other subsections establish different MRM/MRW moment approximation theorems  as functions of  $\Omega(t)$ moments. All along these sections, the moment approximation will be made on the following criterium

\begin{Nt}
Let $\{X_{\lambda}(t)\}_t$ and $\{Y_{\lambda}(t)\}_t$ be two 
processes that depend on the parameter $\lambda^2$. Let $M_{X_\lambda}(t_1,...t_n)$ be a given generalized moment of the process
$\{X_{\lambda}(t)\}_t$. Let $M_{Y_\lambda}(t_1,...t_n)$, the corresponding generalized moment of the process $\{Y_{\lambda}(t)\}_t$.
Let us consider the Taylor series (for $\lambda^2$ around 0) of these moments. In the case the zero orders as well as the first following non trivial orders of these Taylor series are identical for any finite generalized moment, we will write
\begin{equation}
\label{notation}
X_{\lambda}(t)\overset{\lambda}{\simeq}Y_{\lambda}(t).
\end{equation}
\end{Nt}

\subsection{The renormalized magnitude $\Omega(t)$}
\label{ssection:renmag}
Let us define the process $\Omega(t)$ that is at the heart of our approximation theory:  
let $\omega_{l,T}(t)$ be the Gaussian process defined in 
Eqs (\ref{def:mean}) and (\ref{def:magcov}) or Eq. (\ref{def:mag}). We define 
the Gaussian process $\Omega_l(t)$ as
\begin{equation}
\label{def:Omlt}
\Omega_l(t)=\frac{1}{\lambda}\int\limits_{0}^{t}
\big(\omega_{l,T}(s)-\EE{\omega_l}\big)ds,
\end{equation}
The renormalized magnitude process
$\Omega(t)$ is defined as the weak limit of  $\Omega_{l}(t)$ : 
\begin{Th}
The process $\{\Omega_{l}(t)\}_{t}$ admits a weak limit when $l$ goes to 0:
\begin{equation}
\label{lim:Omlt}
\Omega(t)=\lim_{l\to 0^{+}}\Omega_l(t)
\end{equation}
\end{Th}
 \begin{IEEEproof}
 The proof of this theorem is a direct consequence of Lemma \ref{lm:conloifin} (Convergence of the finite dimensional laws) and Lemma \ref{lm:tightness} (tightness) of Appendix \ref{app:renmag}.
 \end{IEEEproof}
 
In the sequel, if $I=[t-\tau,t]$ is some interval,
$\Omega(I)$ will stand for the variation of the renormalized 
magnitude over this interval: 
\begin{equation}
\label{eqn:omegai}
\Omega(I) = \delta_{\tau}\Omega(t) = \Omega(t)-\Omega(t-\tau).
\end{equation}

The exact expression of the covariance of the renormalized magnitude
can be simply computed using Lemma \ref{lm:conloifin} of Appendix \ref{app:renmag}:
\begin{Pp}
\label{prop:covmag}
Let $\tau>0$ and $h \ge \tau$. For all $t$, one has:
\begin{itemize}
\item if $h+\tau\leq T$,
\begin{equation}
\label{eq:clogMRM2}
{\mathbb C}{\mathrm ov}\bigg[\frac{\delta_{\tau}\Omega(t)}{\tau},
\frac{\delta_{\tau}\Omega(t+h)}{\tau}\bigg]
=\ln\bigg(\frac{Te^{3/2}}{h}\bigg)+f\left(\frac h \tau \right),
\end{equation}
where the function $f(u)$ reads
\begin{equation}
\label{eq:clogMRM2b}
f(u)=
\begin{cases}
-\frac{(u+1)^2}{2}\ln\big(1+\frac{1}{u}\big)
-\frac{(u-1)^2}{2}\ln\big(1-\frac{1}{u}\big),&\text{if $u\geq 2$},
\\
-2\ln(2),&\text{if $u=1$},
\\
0,&\text{if $u=0$},
\end{cases}
\end{equation}
\item if $h\geq T+\tau$,
\begin{equation}
\label{eq:clogMRM2bis}
{\mathbb C}{\mathrm ov}\big[\delta_{\tau}\Omega(t),
\delta_{\tau}\Omega(t+h)\big]=0.
\end{equation}
\end{itemize}
\end{Pp}
Let us note that in the case $\tau \ll  h <T+\tau$, Eqs \eqref{eq:clogMRM2} and \eqref{eq:clogMRM2b} simplify a lot. Indeed, in this case,  the function $f(u)$ 
becomes $f(u) =-3/2 +  {\mathcal O}(1/u)$, one thus gets
\begin{Cr}
\label{cr:covmag}
Let $\tau\ll h<T+\tau$, then  for all $t$, one has:
\begin{equation}
\label{coro}
{\mathbb C}{\mathrm ov}\bigg[\frac{\delta_{\tau}\Omega(t)}{\tau},
\frac{\delta_{\tau}\Omega(t+h)}{\tau}\bigg]
=\ln\bigg(\frac{T}{h}\bigg)+{\mathcal O}(\tau/h),
\end{equation}
\end{Cr}

We are now ready to formulate the main approximation results one can obtain
in the limit of small intermittency $\lambda^2 \rightarrow 0$.

\subsection{Convergence in law towards the renormalized magnitude}
\label{ssec:law}
One can prove an asymptotic theorem concerning the logarithm of the measure of
an interval. More precisely, one has the following result:
\begin{Th}
\label{thm:renmagconv}
Let  $I_1,\ldots,I_n$, be $n$ arbitrary intervals. When $\lambda^2$ goes to zero 
we have the following convergence in law:
\begin{equation}
\bigg(\frac{1}{2\lambda}\ln\bigg(\frac{M(I_1)}{|I_1|}\bigg),
\ldots,\frac{1}{2\lambda}\ln\bigg(\frac{M(I_n)}{|I_n|}\bigg)\bigg)
\overset{{\mathcal L}}{\longrightarrow} \bigg(\frac{\Omega(I_1)}
{|I_1|},\ldots,\frac{\Omega(I_n)}{ |I_n|}\bigg).
\end{equation} 
\end{Th}

\begin{IEEEproof}
From Proposition \ref{thm:momcgen} of Appendix \ref{appsi} and Proposition \ref{prop:momloggen} of Appendix \ref{app:log}, it results that,  
for all $n$,
\begin{equation}
\lim_{\lambda\rightarrow 0}\EE{\prod_{j=1}^{n}\frac{1}{2\lambda} 
\ln\bigg(\frac{M(I_j)}{|I_j|}\bigg)} = \EE{\frac{\Omega(I_1)}
{|I_1|},\ldots,\frac{\Omega(I_n)}{ |I_n|}}.
\end{equation}
A simple multidimensional generalization of the Theorem 4.5.5 in \cite{Chu74} allows one to deduce the convergence in law from the convergence of the generalized moments \cite{Koz06}.
\end{IEEEproof}

The following corollary on the successive incremements of the measure is
a direct consequence of the previous theorem: 
\begin{Cr}
\label{cor:conv2}
If $\tau > 0$, then
\begin{equation}
\bigg\{\frac{1}{2\lambda}\ln\bigg(\frac{\delta_{\tau}M(t)}{\tau}\bigg)\bigg\}_{t}
\overset{{\mathcal L}}{\longrightarrow} \bigg\{\frac{\delta_{\tau}\Omega(t)}{\tau}\bigg\}_{t}.
\end{equation}
\end{Cr}

\subsection{Approximation of the moments of the logarithm of the measure}
\label{ssec:amlm}
The following result will be particularly useful for the estimation of log-normal MRM as 
discussed in section \ref{sec:est} below. 

\begin{Th}[Convergence of the magnitude generalized moments]
\label{thm:MRMlogn1}
At scale $\tau>0$ the process 
$\{2\lambda\delta_{\tau}\Omega(t)/\tau\}_{t}$ reproduces 
the Taylor series (in $\lambda^2$), up to the first non trivial 
order,  of any finite generalized moment of the logarithm of the log-normal MRM increments (see Eq. \eqref{notation} for precision on 
the following notation):
\begin{equation}
\label{eq:MRMlogn1}
\ln\left(\frac{\delta_{\tau}M(t)}{\tau}\right)\overset{\lambda}{\simeq} 
2\lambda\frac{\delta_{\tau}\Omega(t)}{\tau}.
\end{equation}
\end{Th}

\begin{IEEEproof}
This result is a direct consequence 
of Proposition \ref{thm:momcgen} of Appendix \ref{appsi} and Proposition \ref{prop:momloggen} of Appendix \ref{app:log}.
\end{IEEEproof}

This theorem allows one to obtain approximations
of the mean and of the covariance function of logarithm of the MRM increments:

\begin{Th}[Magnitude mean and covariance approximations]
\label{th:covM}
For all $\tau>0$ and $h\geq 0$ and $t$, one has
\begin{equation}
\label{eq:covM}
{\mathbb C}{\mathrm ov}\bigg[\ln\bigg(\frac{\delta_{\tau}M(t)}{\tau}\bigg),
\ln\bigg(\frac{\delta_{\tau}M(t+h)}{\tau}\bigg)\bigg]
=4\lambda^2{\mathbb C}{\mathrm ov}\bigg[\frac{\delta_{\tau}\Omega(t)}{\tau},
\frac{\delta_{\tau}\Omega(t+h)}{\tau}\bigg]+o(\lambda^2),
\end{equation}
where the covariance of the increments of the renormalized magnitude 
is provided by Lemma \ref{lm:conloifin} in Appendix \ref{app:renmag}. Moreover, as far as $h+\tau\leq T$, then 
the term $o(\lambda^2)$ no longer depends on the integral scale $T$ and depends on 
$\tau$ only through the ratio $h/\tau$ and goes to $0$ when $\tau \rightarrow 0$ (with $h$ fixed).
Moreover
\begin{equation}
\label{eq:mlogMRM}
\EE{\ln\left(\frac{\delta_{\tau}M(t)}{\tau}\right)}
=-2\lambda^2 \ln\left(\frac{Te^{3/2}}{\tau}\right)+o(\lambda^2),
\end{equation}
where the term $o(\lambda^2)$ depends neither on $T$ nor on $\tau$ 
\end{Th}

\begin{IEEEproof}
The relationship \eqref{eq:covM} directly results from previous theorem. 
We simply have to show that for $h+\tau\leq T$, $o(\lambda^2)$ depends only on $h/\tau$ and goes to 0 when $\tau \rightarrow 0$.

By using the invariance properties (\ref{eq:gscaleinv}) and  (\ref{eq:iscaleinv}) of the Proposition \ref{prop:scaleinv}
we get the equality in law:
\begin{equation}
\label{eq:bb}
\big\{\delta_{\tau}M_{T}(t)\big\}_{\tau\leq t\leq h}
\overset{{\mathcal L}}{=}W_{h/T}\big\{ \delta_{\tau}M_{h/\tau}(t/\tau)\big\}_{\tau \leq t\leq h},
\end{equation}
where $W_{h/T}$ is a log-normal random variable that satisfies
\begin{equation}
\label{eq:prec}
{\mathbb V}{\mathrm ar}\big[\ln(W_{h/T})\big]=4\lambda^2\ln\left(\frac{T}{h}\right).
\end{equation}
From \eqref{eq:bb}, one can easily prove that the 
difference  ${\mathbb C}{\mathrm ov}\big[\ln(|\delta_{\tau}M_{T}(t)|),
\ln(|\delta_{\tau}M_{T}(t+h)|)\big]-{\mathbb V}{\mathrm ar}\big[\ln(W_{h/T})\big]$ 
depends only on $\lambda^2$ and  $h/\tau$. The fact that it goes to 0 when $\tau$ goes to 0 comes from a straightforward argument using the cone representation.
Moreover, thanks to Lemma \ref{lm:conloifin} in Appendix \ref{app:renmag},
one obtains, if $h+\tau\leq T$,
\begin{equation}
4\lambda^2{\mathbb C}{\mathrm ov}\bigg[\frac{\delta_{\tau}\Omega(t)}{\tau},
\frac{\delta_{\tau}\Omega(t+h)}{\tau}\bigg]
\!=\!4\lambda^2\!\int\limits_{t-\tau}^{t}\!\frac{du}{\tau}\!\int\limits_{t+h-\tau}^{t+h}\!\frac{dv}{\tau}\!
\ln\bigg(\frac{h}{|u-v|}\bigg)+{\mathbb V}{\mathrm ar}\big[\ln(W_{h/T})\big].
\end{equation}
By choosing the new variables $u'=u/\tau$ and $v'=v/\tau$, we can show that the above integral 
depends only on $h/\tau$ and goes to 0 when $\tau \rightarrow 0$ (with $h$ fixed). By inserting this expression in 
Eq.  \eqref{eq:covM}, we thus conclude that the terms $o(\lambda^2)$ in this equation 
depends only on $h/\tau$.

A similar computation allows us to deduce Eq. (\ref{eq:mlogMRM}).
\end{IEEEproof}

\subsection{Approximation of the moments of  the measure}
\label{ssec:amm}
As far as the generalized moments of the measure itself ($\delta_{\tau}M(t)$)
are concerned, Theorem \ref{thm:MRMlogn1}
suggests that they could be well reproduced to the first non trivial order
by the moments of the process 
$\tau e^{2\lambda\frac{\delta_{\tau}\Omega(t)}{\tau}}$. It is easy to
see that this cannot be true. Indeed, the mean of the two previous processes
are different simply because the expectation of the exponential of a random
variable is not the exponential of its expectation.
It is therefore necessary to slightly modify the process
$\frac{\delta_{\tau}\Omega(t)}{\tau}$ by changing its mean value. 

\begin{Th}
\label{thm:MRMlogn2}
Let  $\tau > 0$. The process 
$\big\{\tau e^{2\lambda\delta_{\tau}\Omega(t)/\tau
-2\lambda^2{\mathbb V}{\mathrm ar}[\delta_{\tau}\Omega(t)/\tau]}\big\}_{t}$ 
reproduces 
the Taylor series (in $\lambda^2$), up to the first non trivial 
order,  of any finite generalized moment of the log-normal MRM increments (see Eq. \eqref{notation} for precision on 
the following notation):
\begin{equation}
\label{eq:MRMlogn2}
\delta_{\tau}M(t)\overset{\lambda}{\simeq}\tau e^{{2\lambda}\frac{\delta_{\tau}\Omega(t)}{\tau}
-2\lambda^2{\mathbb V}{\mathrm ar}\big[\frac{\delta_{\tau}\Omega(t)}{\tau}\big]}.
\end{equation}
\end{Th}
\begin{IEEEproof}
The $n$-points moment of the r.h.s. process can be written as:
\begin{equation}
m(t_{1},\ldots,t_{n})
=\tau^n e^{-2n\lambda^2{\mathbb V}{\mathrm ar}\big[\frac{\delta_{\tau}\Omega}{\tau}\big]}
\EE{e^{2\lambda\sum_{i=1}^{n}\frac{\delta_{\tau}\Omega(t_{i})}{\tau}}}
=\tau^n e^{2\lambda^2 {\mathbb V}{\mathrm ar}\big[\sum_{i=1}^{n}\frac{\delta_{\tau}\Omega(t_{i})}{\tau}\big]}.
\end{equation}
If one considers the Taylor series expansion of this expression and replaces
the variance of $\sum_{i=1}^{n}\delta_{\tau}\Omega(t_{i})$ 
by its expression (provided by Lemma \ref{lm:conloifin} in Appendix \ref{app:renmag}), one gets:
\begin{equation}
m(t_{1},\ldots,t_{n})=\tau^n
+4\tau^n\int\limits_{t_{n}}^{t_{n}+\tau}{du_1}\cdots\int\limits_{t_{n}}^{t_{n}+\tau}{du_n}
\sum_{1\leq j\leq k\leq n}\rho(u_{j}-u_{k})
+o(\lambda^2),
\end{equation}
where $\rho$ is defined by \eqref{eq:rho}. 
Using Lemma \ref{lm:fl} in Appendix \ref{appsi}, it follows 
\begin{equation}
m(t_{1},\ldots,t_{n})
=\EE{\delta_{\tau}M(t_{1}),\ldots,\delta_{\tau}M(t_{n})}+o(\lambda^2),
\end{equation}
which leads to the expected result.
\end{IEEEproof}

\subsection{Approximation of the MRW process}
\label{sec:appMRWln}

The log-normal MRW process being defined by a Brownian motion subordinated with a log-normal MRM measure,
it is obvious that the generalized moments of its increments and their logarithm are related to
those of the MRM measure. In previous sections we have obtained an approximation of these MRM
generalized moments. The Theorems \ref{thm:MRMlogn1}, \ref{th:covM}  and \ref{thm:MRMlogn2} naturally extend to MRW increments. The following theorems are direct consequences from these theorems.

\begin{Th}
\label{thm:MRWlogn2}
Let $\tau>0$ and $\{\epsilon[n]\}_{n}$ a gaussian white noise of variance  
$\sigma^2$. \\ The discrete time process  
$\big\{\tau^{1/2}\epsilon[n]e^{\delta_{\tau}\Omega(n\tau)/\tau
-\lambda^2{\mathbb V}{\mathrm ar}[\delta_{\tau}\Omega/\tau]}\big\}_{n}$ 
reproduces the Taylor series (in $\lambda^2$), up to the first non trivial 
order,  of any finite generalized moment of  the increments of a MRW process $X(t)$ (see Eq. \eqref{notation} for precision on 
the following notation):
\begin{equation}
\label{eqn:MRWlogn2}
\delta_{\tau}X(n\tau)\overset{\lambda}{\simeq}
\tau^{1/2}\epsilon[n]e^{{\lambda}\frac{\delta_{\tau}\Omega(n\tau)}{\tau}
-\lambda^2{\mathbb V}{\mathrm ar}\big[\frac{\delta_{\tau}\Omega}{\tau}\big]}.
\end{equation}
Moreover the first non trivial order is of order $\lambda^2$.
\end{Th}

\begin{Th}
\label{thm:MRWlogn1}
Let $\tau > 0$ and $\{\epsilon[n]\}_{n}$ a gaussian white noise of variance 
$\sigma^2$. \\ The discrete time process $\big\{\ln(\tau^{1/2})+\ln(|\epsilon[n]|)
+\lambda \delta_{\tau}\Omega(n\tau)/\tau\big\}_{n}$ 
reproduces the Taylor series (in $\lambda^2$), up to the first non trivial 
order,  of any finite generalized moment of the absolute increments of a MRW process $X(t)$ (see Eq. \eqref{notation} for precision on 
the following notation):
\begin{equation}
\label{eq:MRWlogn1}
\ln\left|\delta_{\tau}X(n\tau)\right|
\overset{\lambda}{\simeq}\frac{1}{2}\ln(\tau)+\ln(|\epsilon[n]|)
+\lambda\frac{\delta_{\tau}\Omega(n\tau)}{\tau}.
\end{equation}
As in previous theorem, the first non trivial order is of order $\lambda^2$.
\end{Th}

\begin{Th}
\label{th:covM_mrw}
For all $\tau>0$ et $h\geq 0$, one has, for all $t$
\begin{equation}
\label{eq:covX}
R_{\tau}(h) = {\mathbb C}{\mathrm ov}\big[\ln(|\delta_{\tau}X(t)|),\ln(|\delta_{\tau}X(t+h)|)\big]
=\frac{\pi^2}{8}\delta(h)+\lambda^2{\mathbb C}{\mathrm ov}\bigg[\frac{\delta_{\tau}\Omega(t)}{\tau},
\frac{\delta_{\tau}\Omega(t+h)}{\tau}\bigg]+o(\lambda^2),
\end{equation}
where the covariance of the increments of the renormalized magnitude 
is provided by Lemma \ref{lm:conloifin} in Appendix \ref{app:renmag}. Moreover, as far as $h+\tau\leq T$, then 
the term $o(\lambda^2)$ no longer depends on the integral scale $T$ and depends on 
$\tau$ only through the ratio $h/\tau$ and goes to 0 when $\tau \rightarrow 0$ (with $h$ fixed). 
Moreover, one has
\begin{equation}
\label{eq:3}
\mathbb{E}\big[\ln(|\delta_{\tau}X(n\tau)|)\big]=-\frac{\gamma+\ln(2)}{2}
-\lambda^2\ln\bigg(\frac{Te^{3/2}}{\tau}\bigg)+o(\lambda^2),
\end{equation}
where the term $o(\lambda^2)$ depends neither on $T$ nor on $\tau$ 
and 
$\gamma$ is the Euler constant.
\end{Th}
In the case  $\tau \ll  h <T+\tau$, Eq. \eqref{coro} gives an approximation of the auto covariance of the renormalized magnitude. It can be used to get an approximation of the auto covariance $R_{\tau}(h)$, i.e.,
\begin{Cr}
\label{cr:covX}
Let $\tau\ll h<T+\tau$, then  for all $t$, one has:
\begin{equation}
\label{eq:crcovX}
R_{\tau}(h) = {\mathbb C}{\mathrm ov}\big[\ln(|\delta_{\tau}X(t)|),\ln(|\delta_{\tau}X(t+h)|)\big] = \lambda^2 \ln\bigg(\frac{T}{h}\bigg)+ \lambda^2 {\mathcal O}(\tau/h) + o(\lambda^2).
\end{equation}
where the term $o(\lambda^2)$ no longer depends on the integral scale $T$ and depends on 
$\tau$ only through the ratio $h/\tau$ and goes to 0 when $\tau \rightarrow 0$ ($h$ fixed). 
\end{Cr}

\section{Parameter estimation}
\label{sec:est}

We have seen that a log-normal cascade model is mainly defined by 
2 parameters (apart from the variance parameter $\sigma^2$ which is a simple multiplicative factor for the MRW):  the integral scale $T$ and the intermittency coefficient $\lambda^2$.
Among the huge literature devoted to multifractal models and multifractal analysis,
there are only very few papers that focus on issues related to parameter estimation or related 
statistical questions (see however \cite{OssWay00, ResSamGilWil03, Lux03, Lux04, CalFis04}).   

A simple method to estimate $\lambda^2$ would consist in performing a
regression of the empirical $\zeta(q)$ function estimated from the scaling 
behavior of the empirical moments. However this method is far from being robust, the variance of this estimator converges very slowly (of the type $N^{-1+\alpha}$ with $\alpha > 0$, see \cite{OssWay00}).
This method is however sufficient to establish the pertinence of the 
approximation $\lambda^2\ll 1$ in many empirical situations like the analysis of turbulence 
or financial time series \cite{MuzDelBac00,BacDelMuz01}.
The starting point of our approach of parameter estimation is therefore to 
assume that we are in the small intermittency regime $\lambda^2 \ll  1$ and that the results of section \ref{sil} 
can be used. 

Let $N = \frac L \tau $ be the total number of samples available, where $L$ is the observation scale and $\tau$ the sampling period.
Consequently, the observed samples corresponds to the values
\begin{equation}
\{X_T(n\tau)\}_{n\in [0,N[},~~\mbox{where~} N = \frac L \tau.
\end{equation}
 The estimation problem must be studied in
the asymptotic regime $N \rightarrow +\infty$. However, this limit
can be achieved in two different ways. The first one, referred to as {\em low-frequency regime}, 
corresponds to the case where $\tau$ is fixed and  $L \rightarrow + \infty$. 
In the second one, referred to as {\em high frequency regime},  $L$ is fixed but the 
and $\tau\rightarrow 0$. From a numerical point of view, $L\gg T$, corresponds to the low-frequency regime whereas  $\tau\ll T$ corresponds to the high-frequency regime. In both cases $N = L/\tau \rightarrow +\infty$. In the particular case where one has both $L\gg T\gg \tau$, the effective asymptotic can be considered to be the high 
(resp. low) frequency regime if $\frac L T \ll  \frac T \tau$ (resp. $\frac L T \gg  \frac T \tau$). For discussions on 
{\em mixed regime} for which $L\rightarrow +\infty$ and $\tau \rightarrow 0$ at the same time, we refer the reader to \cite{MuzBacKoz06, BacKozMuz06,MuzBacBaiPog08, BacGlotHofMuz08}.

\subsection{GMM  in the low frequency regime, $L\rightarrow +\infty$ }
\label{sec:GMM}
The first application of GMM  to estimate multifractal models 
can be found in econometric literature. More precisely,
Calvet and Fisher \cite{CalFis01,CalFis04} used this method to estimate
the parameters of a simple cascade model where the random weights
follow a binomial law. Their work has been further developed by Lux \cite{ Lux03, Lux04}.

It is easy to see that the three parameters $\lambda^2$, $T$ and $\sigma^2$,
are directly related to some moments associated with MRW increments or their logarithm.
It is therefore natural to use a GMM  to estimate these parameters.
GMM was initially proposed by Hansen \cite{Han82} and can be described as follows:

Let us consider the process $\{Z_\tau^{(\theta)}[k]\}_k$ of the logarithms of absolute increments of some MRW process at
size $\tau$  :
\begin{equation}
Z_\tau^{(\theta)}[k] = \ln |\delta_{\tau}X[k]|.
\end{equation}
 This process is characterized by $p=3$ parameters : 
\begin{equation}
  \theta = \left\{ \ln(\sigma),\lambda^2,\ln T \right\}.
\end{equation}
Given some observation $\{Z_\tau^{(\theta_0)}[k]\}_k$,
let us denote $f(Z_\tau^{(\theta_0)}[k],\theta)$ the {\em moment function} of dimension $r>p$, 
which satisfies the following {\em moment condition}:
\begin{equation}
\label{eq:conmom}
\EE{f(Z_\tau^{(\theta_0)}[k],\theta)}=0,~\text{if and only if $\theta=\theta_0$}.
\end{equation}
In our case, 
 it is  natural to choose
the variance of the process $\{e^{2Z_\tau^{(\theta_0)}[k]}\}_k$ in order
to estimate $\sigma^2$ and the empirical covariance of $Z_\tau^{(\theta_0)}$
at various time lags in order to estimate $\lambda^2$ and $T$.
This leads us to consider
\begin{equation}
\label{eq:gN}
f(Z_\tau^{(\theta_0)}[k],\theta)=\left(
\begin{array}{c}
\exp(2Z_\tau^{(\theta_0)}[k]
\\
\big(Z_\tau^{(\theta_0)}[k]-\mu_{\theta}\big)\big(Z_\tau^{(\theta_0)}[k-h_1]-\mu_{\theta}\big)
\\
\vdots
\\
\big(Z_\tau^{(\theta_0)}[k]-\mu_{\theta}\big)\big(|Z_\tau^{(\theta_0)}[k-h_K]-\mu_{\theta}\big)
\end{array}
\right)
-
\left(\begin{array}{c}
\sigma^2\tau
\\
C_{\theta}[h_1]
\\
\vdots
\\
C_{\theta}[h_K]
\end{array}\right),
\end{equation}
where
\begin{equation}
\label{eq:mutheta}
\mu_{\theta}=\EE{Z_{\tau}[k]}
\end{equation}
and
\begin{equation}
\label{eq:Ctheta}
C_ \theta[h]={\mathbb C}{\mathrm ov}\big[Z_\tau^{(\theta)}[k],Z_\tau^{(\theta)}[k-h]\big],
\end{equation}
and $h_1,\ldots,h_K$ are $K$ different positive lags. Let us note that, a first order (in $\lambda^2$) analytical expression of $C_{\theta}[h] = R_{\tau}(h\tau)$ is provided by Eq. \eqref{eq:covX}.

The moment condition \eqref{eq:conmom} can be approximated by using the empirical mean:
\begin{equation}
g_N(\theta)=\frac{1}{N}\sum_{k=1}^{N}f(Z_\tau^{(\theta_0)}[k],\theta).
\end{equation}

The GMM estimator is then simply defined by
\begin{equation}
\label{eq:estGMM}
{\widehat \theta}= argmin_{\theta} \big( g_N^T W_N g_N\big),
\end{equation}
where $W_N$ is a sequence of weighting matrices that converges, when $N \rightarrow +\infty$ towards some matrix positive definite $W_{\infty}$. 
Hansen has established the following result:

\begin{Th}[Hansen \cite{Han82}]
\label{th:gmmth}
If the following hypotheses hold:
\begin{itemize}
\item The process $\{Z_\tau^{(\theta_0)}[k]\}_k$ is ergodic,
\item The series $\{f(Z_\tau^{(\theta_0)}[k],\theta)\}_k$ satisfies a central limit theorem, i.e., 
\begin{equation}
\frac{1}{\sqrt{N}}\sum_{k=1}^{N}f(Z_\tau^{(\theta_0)}[k],\theta)\to{\mathcal N}(0,V_{\theta}),
\end{equation}
where the matrix $V_{\theta}$ is defined as:
\begin{equation}
\label{eq:V}
V_{\theta}=\lim_{M\to+\infty}\sum_{k=-M}^{M}
\EE{f(Z_\tau^{(\theta_0)}[k],\theta)f(Z_\tau^{(\theta_0)}[k],\theta)^T}.
\end{equation}
\item The $(r\times p)$ matrix $Dg_N=\frac{\partial g_N}{\partial\theta}$ 
has full rank ($p$) and converges towards 
\begin{equation}
Df=\EE{\frac{\partial f(Z_\tau^{(\theta_0)}[k],\theta)}{\partial\theta}},
\end{equation}
\end{itemize}
then, the GMM estimator $\widehat{\theta}$ is consistent and verifies
\begin{equation}
\label{eq:convloi}
\sqrt{N}\big(\widehat{\theta}-\theta\big)\to\mathcal{N}(0,\Sigma),
\end{equation}
where
\begin{equation}
\Sigma=\big(Df^T W_{\infty}Df\big)^{-1}Df^T W_{\infty} V_{\theta_0} 
W_{\infty}Df\big(Df^T W_{\infty}Df\big)^{-1}.
\end{equation}
Moreover, the estimator $\widehat{\theta}$ is optimal if $W_{\infty}=V_{\theta_0}^{-1}$, as defined in Eq. \eqref{eq:V}.  In that case
the asymptotic covariance of the estimator is 
\begin{equation}
\Sigma_{opt}=\big(Df^T V_{\theta_0}^{-1}Df\big)^{-1}.
\end{equation}
\end{Th}
In practice \cite{Hall05}, it is obviously difficult to use the optimal weighting matrix $W_{\infty}=V_{\theta_0}^{-1}$
since one does not know the vector $\theta_0$. One usually proceeds using the following
iterative algorithm:
\begin{enumerate}
\item Choose some arbitrary initial weighting matrix $W_N$, such as $\mathbb{I}\mathrm{d}_N$,
\item Compute the GMM estimator \eqref{eq:estGMM} using this matrix $W_N$,
\item Replace the  weighting matrix by $W_N=V_{\widehat{\theta}}^{-1}$, where
$\widehat{\theta}$ is the obtained estimated parameter vector.
\item Repeats step 2 and 3 until successive estimates are sufficiently close one to each other.
\end{enumerate}
Confidence intervals for $\widehat{\theta}$ can be obtained using Eq. \eqref{eq:convloi}.

One can easily show that the hypothesis of Theorem \ref{th:gmmth} hold in the case the moment function is defined by \eqref{eq:gN}.
However, there is one major problem for implementing the corresponding GMM method : we do not have any analytical expressions neither of $\mu_{\theta}$ (Eq. \eqref{eq:mutheta})
nor $C_ \theta[h]$ (Eq. \eqref{eq:Ctheta}), nor $V_{\theta}$ (Eq. \eqref{eq:V}) . Actually Eqs \eqref{eq:covX} and \eqref{eq:3} of Theorem \ref{th:covM_mrw} give analytical approximations (up to a $o(\lambda^2)$ term) to both $\mu_{\theta}$ and $C_ \theta[h]$. Let us note that these very same equations also allow to derive an analytical expression (up to a $o(\lambda^2)$ term) of $V_{\theta}$ \cite{Koz06}.
It is tempting to use these approximations in the moment function \eqref{eq:gN} and for the weighting matrix
 and try to use the exact same GMM algorithm. This is exactly the framework of the so-called {\em GMM estimation in a misspecified model} (see \cite{Hall05}). The model is considered as {\em misspecified} since the moment function no longer satisfies \eqref{eq:conmom}. Instead, one has
 \begin{equation}
 \label{eq:conmommis}
 \EE{f^*(Z_\tau^{(\theta_0)}[k],\theta)}=r(\theta),~\text{for all}~k,~\text{and}~||r(\theta)|| > 0~\text{for all}~\theta,
 \end{equation}
 where $f^*$ corresponds to the moment function \eqref{eq:gN} in which we have substituted $\mu_{\theta}$ and $C_ \theta[h]$ by their approximations.
Now,  if we suppose that there exists $\theta^*$ such that
\begin{equation}
\EE{f^*(Z_{\theta_{0}},\theta^*}^T W_{\infty} \EE{f^*(Z_{\theta_{0}},\theta^*} < \EE{f^*(Z_{\theta_{0}},\theta}^T W_{\infty}  \EE{f^*(Z_{\theta_{0}},\theta},~~~
\forall {\theta}\neq {\theta}^*,
\end{equation}
using the results of \cite{Hall05}, one can show that the so-obtained {\em approximated} GMM gives a consistent asymptotically gaussian estimator of $\theta^*$.  Moreover, we expect
\begin{equation}
\label{thetaapprox}
\theta^* = \theta + o(\lambda^2).
\end{equation}
In order, to illustrate this estimation method, we have run a Monte-Carlo test on MRW realizations. The results are shown in Table \ref{GMMresults}.
Each MRW was simulated on a discrete time grid of period $\tau = 1$ and of various size $L$ using the algorithm described in Section \ref{montecarlo}\footnote{As explained in this section, we chose $l=128$. Let us note that increasing $l$ does not significantly change the numerical results}. The number $K$ of different lags $h_{k}$ used in the moment function \eqref{eq:gN} is $K=43$ and the lags $h_{k}$ are approximately logarithmically distributed between 1 and 150.

For each set of parameters, we simulated 10000 realizations of such MRW and ran the misspecified GMM algorithm on each of these realizations. For each parameter ($\ln(\sigma)$, $\lambda^2$, $\ln(T)$) we computed associated GMM estimators
($\widehat{\ln(\sigma)}$, $\widehat{\lambda^2}$, $\widehat{\ln(T)}$). 
We then computed for each of them the so-obtained bias (the Bias column), the mean square error (MSE column) and we ran the Kolmogorov-Smirnov \cite{ShW86} test for testing the normality of the estimations. The corresponding $p$-values for this test are indicated in the KS column. Thus, for instance, a $5\%$ level test is satisfied if the $p$-value is greater than $0.05$.

Clearly the effective value of $\sigma$ will slightly affect the performance of the GMM algorithm, since it just corresponds to a multiplicative factor. Thus, in all the numerical experiments  we arbitrarily set it to $\sigma = 1$ (i.e., $\ln(\sigma) = 0$). 
The global scale invariance property \eqref{eq:gscaleinv} shows that changing the value of the parameter $T$ amounts to changing the number of samples $L/\tau = L$ (since $\tau = 1$)  
of the realizations. Consequently, the realizations only depend on the ratio $L/T$, i.e., the number of integral scales in a realization. We arbitrarily choose  to 
fix $T$ and have $L$ varying.
In this section, we only adress the low frequency regime 
$T\ll  L$. We  choose $T=200$ (i.e., $\ln(T) \simeq 5.298...$) and $L$ among $\{2048,4096,8192,16384,65536\}$, i.e., the number of integral scales $L/T$  varies from 10 to more than 320. We are thus left with only $\lambda^2$ as a ``free'' parameter. We used two different values for $\lambda^2$ : 0.02 and 0.04.
Thus, two different sets of parameters were used : the first set (top half of Table \ref{GMMresults})
corresponds to $\sigma=1$, $\lambda^2 = 0.02$ and  $T=200$ and the second  set (bottom half of Table \ref{GMMresults}) corresponds to 
$\sigma=1$, $\lambda^2 = 0.04$ and $T=200$. Let us note that adding some more lags $h_{k}$ (i.e., increasing $K$) does not significantly improve the results (see the line corresponding to size $L=16384*$ in  Table \ref{GMMresults}
which corresponds to $K=69$ instead of $K=43$).

For all parameters, Table \ref{GMMresults} shows clearly that the MSE is entirely dominated by the variance (the bias contribution is negligeable).
Let us discuss the results obtained for the estimation of  each parameter one after the other.
\begin{itemize}
\item $\widehat{\log{\sigma}}$ :  Clearly, the theoretical GMM asymptotics for the parameter $\ln(\sigma)$ is reached as soon as $N=2048$. This is indicated both by the fact that 
 the MSE decreases as $1/\sqrt{L}$ and that the Kolmogorov-Smirnov normality test has a very high $p$-value as soon as $N\ge2048$ 
 (for $N=2048$, the $p$-value is almost $40\%$ when $\lambda^2=0.02$ and almost  $25\%$ when $\lambda^2=0.04$).  
 \item $\widehat {\lambda^2}$ : For $\lambda^2$ the situation is somewhat different. Though the estimation is surprisingly good even for the shortest realizations in the sense that  the MSE is very small, the normal asymptotics cannot be considered to be reached when $N\le  16384$, i.e., when the number of integral scales $L/T$ is smaller than 80.
 \item $\widehat{\ln T}$ :  Here the GMM asymptotics for parameter $\ln(T)$ is the slowest. Though the MSE is small for $N\ge 16384$, the normal asymptotic can hardly be considered to be reached even for $N=65536$. 
 \end{itemize}
 Let us note that, in any case, the $o(\lambda^2)$ term in \eqref{thetaapprox} due to the mispecification of the model hardly  shows up  in these results. Indeed, we expect a bias of the order of $\lambda^4$ (i.e., the ``next'' order after $\lambda^2$), thus of the order of $4e-04$ for the top half of the Table ($\lambda^2=0.02$) and $1.6 10^{-3}$ for the bottom half ($\lambda^2=0.04$). Except for the case $\lambda^2=0.04$ and for the estimation of 
 the parameter $\lambda^2$ (for which the bias saturates around $2 10^{-5}$),  there does not seem to be any trace of this term : $L$ is not large enough. 
Even when the bias saturates around $2 10{-5}$ (for an MSE of 0.0015), in order this saturation value to dominate the MSE, $L$ should be of the order of 
 $10^{17}$! Thus, though the model is theoretically mispecified, from a practical point of view, it can be considered as well specified.

 As a conclusion to this section, we can state that the GMM estimations are reliable in the low frequency regime, however, except for $\sigma$ the normal asymptotic confidence intervals should not be used. Monte Carlo simulations should be performed to get confidence intervals. 

\begin{table}
\begin{small}
\begin{center}
\begin{tabular}{|c|ccc|ccc|ccc|}
\hline
$\tau = 1$ &\multicolumn{3}{c}{$\ln \sigma = 0$ ($\sigma = 1$)}&\multicolumn{3}{c}{$\lambda^2=0.02$}& \multicolumn{3}{c}{$\ln T\simeq 5.298..$ ($T = 200$)}\\
\cline{2-4}\cline{5-7}\cline{8-10}
$L$ &Bias&MSE&KS& Bias&MSE&KS& Bias&MSE&KS \\
\hline
2048 & -5e-03 & 0.070 & 0.39 &  5e-04 & 0.0072 & 3e-08 & -0.013 & 1.15 & 2e-72 \\
4096 & -2e-03 &  0.049 & 0.49 &  3e-04 & 0.0048 & 1e-03 & -0.026 &  0.76 & 1e-33 \\
8192 & -6e-04 &  0.034 & 0.67 &  1e-04 & 0.0032 & 2e-03 &  -0.015 &  0.50 & 9e-16 \\
16384 & -8e-04 &  0.024 & 0.56 &  2e-05  & 0.0022 & 0.08 & -0.009 &  0.34 & 6e-8 \\
16384* & -9e-04 &  0.024 & 0.54 &  -2e-05  & 0.0022 & 0.08 & 0.005 &  0.35 & 5e-09  \\
65536 & -2e-04 &  0.012 & 0.49 &  6e-06 & 0.0011 & 0.45 & -0.002 &  0.17 & 0.01 \\
\hline \hline
$\tau = 1$ &\multicolumn{3}{c}{$\ln \sigma = 0$ ($\sigma = 1$)}&\multicolumn{3}{c}{$\lambda^2=0.04$}& \multicolumn{3}{c}{$\ln T\simeq 5.298..$ ($T = 200$)}\\
\cline{2-4}\cline{5-7}\cline{8-10}
$L$& Bias&MSE&KS& Bias&MSE&KS& Bias&MSE&KS \\
\hline
2048 & -1e-02 & 0.110 & 0.24 &  7e-04 & 0.0095 & 9e-05 & -0.130 &  0.88 & 5e-32 \\
4096 & -5e-03 & 0.072 & 0.34 &  4e-04 &  0.0064 & 0.01 & -0.054 &  0.59 & 4e-18 \\
8192 & -3e-03 & 0.050 & 0.48 &  2e-05 &  0.0044 & 0.06 & -0.027 &  0.41 & 3e-6 \\
16384 & -2e-03 & 0.035 & 0.52 &  -2e-05 &  0.0031 & 0.08 & -0.014 &  0.28 & 2e-5 \\
65536 & -4e-04 &  0.018 & 0.42 &  -4e-05 &  0.0015 & 0.40 & -0.002 &  0.14 & 0.05 \\
\hline
\end{tabular}
\end{center}
\caption[coucou]
{\label{GMMresults}\small
GMM estimation of MRW parameters. Each line corresponds to GMM estimation as explained in section \ref{sec:GMM} on 10000 realizations of discrete-time MRW with
$\tau = 1$ and of size  $L$. $L$ varies from 2048 to 65536. 
The lags $h_{k}$ used for GMM estimation in Eq. \eqref{eq:gN} are chosen such that $K=43$ and approximately logarithmically distributed between 1 and 150 except for the line corresponding to $N=16384*$ for which more lags were taken 
($K = 69$). 
The MRW  were Monte-Carlo generated using  the algorithm described in Section \ref{montecarlo}. Two sets of  parameters were used : $\sigma=1$, $\lambda^2 = 0.02$, $T=200$ for the top half and $\sigma=1$, $\lambda^2 = 0.04$, $T=200$ for the bottom half.
}

\end{small}
\end{table}

\subsection{GMM estimation in the high frequency regime $\tau \rightarrow 0$ - Estimation  of the nature of the asymptotic regime}
\label{sec:estgrT}

In many practical situations (e.g., when dealing with financial time series) the data are sampled at some high frequency $\tau \ll  T$ over
a time period $L$ that is smaller than (or of the order of) the integral scale $T$. As already explained, in that case, the right asymptotic regime to consider is the high frequency regime $\tau \rightarrow 0$.
Let us try to understand how behaves the previously described GMM procedure in that context.
As we have already pointed out, the moment function \eqref{eq:gN} involved in the GMM has two types of components : 
the first component corresponds to the empirical variance of the  increments of the MRW process itself and basically allows one to estimate $\sigma^2$ while all the other components correspond to the empirical covariance of the logarithm of the same increments and allow one to estimate $T$ and $\lambda^2$. 

According to Eq. \eqref{eq:iscaleinvmrw},
the log-normal MRW process $\{X_T(t)\}_{t\leq L}$ satisfies
the following equality in law: 
\begin{equation}
\big\{X_{T}(t)\big\}_{t\leq L}\overset{{\mathcal L}}{=}
\big\{W_{L/T}X_{L}(t)\big\}_{t\leq L},
\end{equation} 
where $W_{L/T}$ is a log-normal random variable which law is given in Proposition \ref{prop:scaleinvmrw}
and which is independent of the MRW process 
$\{X_{L}(t)\}_{t\leq L}$ which integral scale is equal to the observation 
scale $L$.
Given some sample of length $L$ of the MRW process, the variable $W_{L/T}$
takes a fixed value and can be considered as a simple multiplicative
factor that simply changes the variance of the process. Consequently, the estimation problem of both $\sigma^2$ and $T$ is ill-posed.
It is fundamentally impossible to estimate independently the integral scale $T$ 
and the variance $\sigma^2$ of the process since they both appear as a multiplicative factor of the whole process. $T$ is no longer a ''true'' parameter 
of the model it can be arbitrarily fixed. Moreover, even if we knew the true value of $T$,
there is no chance for the GMM variance estimator $\widehat{\ln \sigma}$  to converge to 0 in the asymptotic limit $\tau \rightarrow 0$ 
since it is easy to show that the empirical variance of the increments, itself,  does converge in the limit  $\tau \rightarrow 0$ towards a random variable (see e.g. \cite{BacMuz03}).
Hence, in this regime, the first hypothesis upon which GMM relies, namely the ergodicity of $\{ Z_{\tau}[n] \}_n$ is not satisfied.

Since the value of $T$ is the key to decide in which asymptotic regime one is ($L\gg T$ for the low frequency regime and $ T\gg \tau$ for the high frequency regime), it is of fundamental interest  to understand how the GMM estimation of $T$ behaves in the high frequency regime $\tau \rightarrow 0$.
Actually, the GMM estimation of $\ln T$  (and of $\lambda^2$ )  basically consists in  fitting  the empirical covariance of the logarithm of the increments of the MRW process. Thus,  it is natural to study the mean of this empirical covariance, in the high frequency regime $\tau \rightarrow 0$.

\begin{Pp}
Let us consider the fixed observation scale $L \le T$ where $\tau$ is the sampling scale and
$L/\tau$ is the number of samples of the MRW process. We introduce the empirical covariance $\hat R_{\tau}[n]$ and the empirical mean $\hat\mu_{\tau,M}$ :
\begin{equation}
\label{empcov}
\hat R_{\tau}[n] = \frac \tau L  \sum_{k=1}^{L/\tau-n-1}\left( \ln|\delta_{\tau}X(k\tau]|-\hat \mu_{\tau,L/\tau}\right)\left( \ln|\delta_{\tau}X((k+n)\tau]|-\hat \mu_{\tau,L/\tau}\right),~~~
\hat \mu_{\tau,M} = \frac 1 M \sum_{k=1}^{M} \ln|\delta_{\tau}X(k\tau]|.
\end{equation}
Let $h>0$, 
then the expectation of the empirical covariance $\hat R_{\tau}[h/\tau]$, in the high frequency asymptotic $\tau \rightarrow 0$, is
\begin{equation}
\label{eq:biaisgti}
\lim_{\tau\rightarrow 0}
\EE{\widehat{R_{\tau}}[h/\tau]}
=\lambda^2\left[\ln\left(\frac{L}{h e^{3/2}}\right)
-\frac{h}{L}\ln\left(\frac{L}{he^{3/2}}\right)
+\frac{h^2}{L^2}\ln\left(\frac{h}{L}\right)
+\frac{(L-h)^2}{2L^2}\ln\left(1-\frac{h}{L}\right)\right]
-\left(1-\frac{h}{L}\right)\frac{\pi^2}{8L}.
\end{equation}
\end{Pp}

\begin{IEEEproof}
One can easily prove the following   general relation that gives the expectation of the empirical correlation function of a given process : 
\begin{equation}
\label{gene}
\EE{\hat R_{\tau}[h/\tau]}-R_{\tau}(h) = - {\mathbb V}{\mathrm ar}\big[\hat \mu_{\tau,L/\tau}\big] - \frac h L \left( 
R_{\tau}(h) + {\mathbb V}{\mathrm ar}\big[\hat \mu_{\tau,L/\tau}\big] - 2 {\mathbb C}{\mathrm ov}\big[\widehat{\mu}_{\tau,h/\tau},\widehat{\mu}_{\tau,L/\tau}\big]
\right)
\end{equation}

According to Eq. \eqref{eq:crcovX} of Corollary \ref{cr:covX}, under the condition $L\leq T$, 
the covariance function $R_{\tau}(h)$ for $h>0$, in the high frequency asymptotic $\tau \rightarrow 0$,  is given by
\begin{equation}
\lim_{\tau\rightarrow 0} R_{\tau}(h)= \lambda^2\ln\left(\frac{T}{h}\right).
\end{equation}

Moreover, from the definition of the empirical mean,
one can write the following equations 
\begin{equation}
\lim_{\tau\rightarrow 0}{\mathbb V}{\mathrm ar}\big[\widehat{\mu}_{\tau,L/\tau}\big]
=\lambda^2\int\limits_0^{L}\frac{du}{L}\int\limits_0^{L}\frac{dv}{L}
\ln\left(\frac{T}{|u-v|}\right)+\frac{\pi^2}{8L}
=\lambda^2\ln\left(\frac{Te^{3/2}}{L}\right)+\frac{\pi^2}{8L},
\end{equation}
and
\begin{multline}
\lim_{\tau\rightarrow 0}{\mathbb C}{\mathrm ov}\big[\widehat{\mu}_{\tau,h/\tau},\widehat{\mu}_{\tau,L/\tau}\big]
=\lambda^2\int\limits_0^{h}\frac{du}{h}\int\limits_0^{L}\frac{dv}{L}
\ln\left(\frac{T}{|u-v|}\right)+\frac{\pi^2}{8L}
\\
=\lambda^2\ln\left(\frac{Te^{3/2}}{L}\right)
-\lambda^2\frac{h}{2L}\ln\left(\frac{h}{L}\right)
+\lambda^2\frac{(L-h)^2}{2Lh}\ln\left(1-\frac{h}{L}\right)+\frac{\pi^2}{8L}.
\end{multline}
Inserting these last three equations in Eq. \eqref{gene} leads to the expected result.
\end{IEEEproof}

Let us remark that $\lim_{\tau\rightarrow 0}
\EE{\widehat{R_{\tau}}[h/\tau]}$
 does not depend on the integral scale $T$. This is not surprising considering the remark we just made at the beginning of this section. Now, the leading
term of Eq.  \eqref{eq:biaisgti}, when $L \gg  h$  is
\begin{equation}
\label{apppp}
\lim_{\tau\rightarrow 0}
\EE{\widehat{R_{\tau}}[h/\tau]}
\simeq \lambda^2\ln\bigg(\frac{Le^{-3/2}}{h}\bigg).
\end{equation}
Identifying this equation with Eq. \eqref{eq:crcovX}, shows that we expect 
\begin{itemize}
\item[(i)] the estimator $\widehat{\lambda^2}$ to be unbiased and 
\item[(ii)] the mean of the estimator $\widehat{\ln T}$ to be of the order 
of $\EE{\widehat{\ln T}} \simeq \ln(Le^{-3/2}) = \ln(L) -3/2$, independently of the ``true'' integral scale $T$ value. 
\end{itemize}
These results are  illlustrated in Table \ref{GMMresults1} which 
displays the output of the GMM estimators described in the previous section. The estimations where computed using a realization of size $L = 8192$ ($\tau = 1$) of a MRW process with parameters 
$\sigma = 1$, $\lambda^2 = 0.02$ and $T= 16384$. The choice $T>L\gg \tau$ clearly corresponds to the high frequency regime.
This table uses the same format as Table \ref{GMMresults} : 
for each parameter ($\ln \sigma$, $\lambda^2$ and $\ln T$), the bias, the mean square error (MSE) and the $p$-value of the Kolmogorov Smirnov normality test (KS) are computed using a Monte-Carlo method with 10000 realizations.

Let us discuss the results obtained for the estimation of  each parameter one after the other.
\begin{itemize}
\item $\widehat{\ln \sigma}$ : As expected, 
the estimator of $\ln \sigma$ has both a very high bias and mean square error (it does not converge!). 
\item $\widehat{\ln T}$ :
the estimator of $\ln T$ is biased, its mean is found to be  $\EE{\widehat{\ln T}} \simeq 9.704-1.98 = 7.724$ which is very close to the expected order $ \ln(L) -3/2 \simeq  7.51$. This can be used as a way to detect the fact that we are in the high frequency regime.
\item $\widehat{\lambda^2}$ : On the contrary,  the estimation of $\lambda^2$ is excellent, the bias and the MSE are of the same order as the ones obtained 
in Table \ref{GMMresults}. In the following section we prove that an estimator of $\lambda^2$ based on the regression of the empirical covariance function of the logarithm of the increments of a MRW is an unbiased and consistent estimator.
\end{itemize}
\begin{table}
\begin{small}
\begin{center}
\begin{tabular}{|c|ccc|ccc|ccc|}
\hline
$\tau = 1$ &\multicolumn{3}{c}{$\ln \sigma = 0$ ($\sigma = 1$)}&\multicolumn{3}{c}{$\lambda^2=0.02$}& \multicolumn{3}{c}{$\ln T\simeq 9.704..$ ($T = 16384$)}\\
\cline{2-4}\cline{5-7}\cline{8-10}
L & Bias&MSE&KS& Bias&MSE&KS& Bias&MSE&KS \\
\hline
8192 & -0.04  & 0.27 & .55 &  1e-04 &  0.003 & 0.26 & -1.98 &  2.44 & 0 \\
\hline
\end{tabular}
\end{center}
\caption[coucou]
{\label{GMMresults1}\small
GMM estimation of MRW parameters in the case $T/\tau \gg  1$ and $L\le T$ . Each line corresponds to GMM estimation as explained in section \ref{sec:GMM} on 10000 realizations of discrete-time MRW with
$\tau = 1$ and of size  $L = 8192$. 
The lags $h_{k}$ used for GMM estimation in Eq. \eqref{eq:gN} are chosen such that $K=43$ and approximately logarithmically distributed between 1 and 150. 
The MRW  were Monte-Carlo generated using the algorithm described in Section \ref{montecarlo}}

\end{small}
\end{table}

\subsection{A convergent estimator of $\lambda^2$ in the high frequency regime $\tau \rightarrow 0$}
\label{sec:estlambda2}
We consider the sequence of the absolute increments of an MRW 
\begin{equation}
Z_\tau[k] = \ln|\delta_\tau X[k]|.
\end{equation}
According to Theorem \ref{th:covM_mrw} and Proposition \ref{prop:covmag}, for any integer $n>0$ such that $n\tau < T$, one has
\begin{equation}
\label{eq:cov1}
R_\tau(n\tau) = {\mathbb C}{\mathrm ov}\big[Z_{\tau}[k],Z_{\tau}[k+n]\big]
= \lambda^2\bigg(\ln\bigg(\frac{Te^{3/2}}{n\tau}\bigg)-f(n)\bigg)+o(\lambda^2),
\end{equation}
where $o(\lambda^2)$ depends only on $\lambda^2$ and $n$ and where the function $f(n)$ is defined by Eq. \eqref{eq:clogMRM2b}.
It follows that, if $n$ and $n'$ are two different integers such that $0<n\tau<T$ and $0<n'\tau<T$, the difference $R_\tau(n\tau)-
R_\tau(n'\tau)$ does not depend neither on the integral 
scale $T$ nor on the sampling scale $\tau$. 
This naturally leads us to the estimation of $\lambda^2$ relying upon 
a simple regression:
\begin{equation}
\label{eq:estcint}
\widehat{\lambda^2}=\frac{\widehat{R}_{\tau}[n]-\widehat{R}_{\tau}[n']}
{g(n)-g(n')},
\end{equation}
where the empirical covariance $\widehat{R}_\tau$ is defined by \eqref{empcov} and the function $g(n)$ by
\begin{equation}
g(n) = f(n)+\ln(n).
\end{equation}
 We then have the following theorem
\begin{Th}
\label{thm:varestl2MRM}
Let $n,n'$ two different integers such that $0<n\tau<T$ and $0<n'\tau<T$. 
In the high frequency asymptotic regime $\tau \rightarrow 0$,
 the estimator defined by \eqref{eq:estcint} is biased with an asymptotic bias of the order of $o(\lambda^2)$. Moreover it is consistent and its variance decreases as
\begin{equation}
\label{eq:majvarMRM}
{\mathbb V}{\mathrm ar}\big[\widehat{\lambda^2}\big]= {\mathcal O}\bigg(\frac{\ln(N)}{N}\bigg),
\end{equation}
where $N = L/ \tau$.
\end{Th}
\begin{IEEEproof}
The proof for the first assertion is straightforward : the estimator \eqref{eq:estcint} has a bias of the form
\begin{equation}
\EE{\widehat{\lambda^2}} -\lambda^2
={\mathcal O}\bigg(\frac{1}{N}\bigg)+o(\lambda^2).
\end{equation}
The hard part of this theorem is to prove the consistency and the speed of convergence. The rigorous proof is tedious and we just give in Appendix \ref{appconsist} 
the main points of this proof, leaving to the reader some uninteresting and long (though straightforward) computations.
\end{IEEEproof}

\section{Application to financial time-series}
\label{apps}

One of the most important problem in finance is the modelling of price fluctuations of a risky asset. 
Since Mandelbrot famous work on the fluctuations of cotton price in early sixties \cite{Man63}, it is well known that 
speculative price variations are poorly described by the standard Geometric Brownian motion 
(see e.g., \cite{Man73a,Man73b,Man73c,Orl99,Wal03}) that does not permit to explain the 
well known intermittent and correlated nature of volatility variations \cite{DinGra96,Man97,BouPot03}.
During the last decade, the availability of huge data sets of high frequency time series has 
permitted intensive statistical studies that lead to uncover a very rich and non trivial 
statistical structure, that is to some degree universal across different assets.
These empirical studies have suggested that financial data share many statistical properties with turbulent velocity 
"intermittent" fluctuations and notably display {\em multiscaling} properties \cite{ManFisCal97,MuzDelBac00,GhaBrePeiTalDod96,BouPot03}. 
In that respect, the phenomenology of multifractal models \cite{ManFisCal97,CalFis02,Lux04,BacKozMuz06} has provided new concepts and tools to analyze 
market fluctuations and in particular the log-normal MRW disussed in this paper has been shown to account very well
the return fluctuations and volatility correlations over a wide range of time horizons. 
In this section we use the previous GMM method to calibrate the MRW model from daily return time series.
We also show that this model provide a simple way to forecast Value at Risk with better performances than classical
GARCH models.

\subsection{The financial time-series}
The financial data we have used in this section are daily ($\tau = 1$) close prices of some french stocks that are part of the CAC40 french index.  This index is computed using 40 of the largest french stocks 
of the euronext market. We only kept  those  with the longest historic. Thus the data consist in the close prices $P(t)$ of 29 stocks between the years 1990 and 2005\footnote{We have adjusted these prices taking into account the dividends and the eventual splits.}. Thus the time-series associated with each stock has $L=3770$ samples. We use here the MRW process as a model for the log of the price $\ln P(t)$.
Since the GMM estimation described in section \ref{sec:GMM} is partly based on the computation of the logarithms of the increments of the MRW, 
in the case the close price does not change from one day to the next, one cannot compute this increment. We chose, in that case, to change the value of the second price randomly by one tick up or down (the tick is the smallest effective change of the price of a stock).

The problem of risk estimation or forecasting  is essential in quantitative finance. However, several risk measures can be used. In this paper, we address two of them, which are widely used : the (historical) volatility and the Value at Risk (VaR). In any case, risk forecasting will make extensive use of  the aggregation properties of section \ref{sil}.
Before explaining how the forecasting is perfomed, one should estimate the MRW parameters.

\subsection{Parameter estimation using GMM}
In this section, we present the results of the GMM estimation as described in section \ref{sec:GMM}. They are sum up in Table \ref{table_cacgmm}.
The first three columns show the estimations of $\ln\sigma$, $\lambda^2$ and $\ln T$ given by the GMM for each stock. And the next two columns show the corresponding values of $\sigma$ and $T$.

\begin{table}
\begin{small}
\begin{center}
\begin{tabular}{lcccccc}
\hline
\hline
&\multicolumn{3}{c}{GMM}&&\multicolumn{2}{c}{}\\
\cline{2-4}\cline{6-7}
Stock name &$\ln(\sigma)$&$\lambda^2$&$\ln(T)$&&$\sigma$&$T~(\text{days})$\\
\hline
Accor&0.711&0.0327&6.792&&2.0363&891\\
Air~Liquide&0.533&0.0157&8.183&&1.7039&3580\\
Alcatel&1.122&0.0157&19.817&&3.0718&40414\\
Axa&0.812&0.0303&7.796&&2.2529&2430\\
Bouygues&0.823&0.0228&8.341&&2.2774&4193\\
Capgemini&1.093&0.0283&8.936&&2.9834&7605\\
Carrefour&0.624&0.0183&8.220&&1.8665&3716\\
Casino~Guichard&0.626&0.0338&5.340&&1.8707&209\\
Danone&0.420&0.0126&8.808&&1.5218&6687\\
Essilor~International&0.684&0.0266&6.384&&1.9822&592\\
L'Or\'eal&0.675&0.0108&9.265&&1.9641&10561\\
Lafarge&0.692&0.0156&7.343&&1.9968&1546\\
Lagardere&0.908&0.0613&6.186&&2.4800&486\\
LVMH&0.715&0.0275&7.602&&2.0440&2003\\
Michelin&0.739&0.0174&6.579&&2.0934&720\\
Pernod Ricard&0.639&0.0169&6.546&&1.8945&697\\
PSA~Peugeot~Citro\"en&0.643&0.0251&5.737&&1.9015&310\\
Pinault~Printemps&0.761&0.0555&6.380&&2.1411&590\\
Publicis&0.895&0.0473&8.317&&2.4484&4092\\
Saint~Gobain&0.720&0.0231&7.153&&2.0554&1278\\
Sanofi-Aventis&0.709&0.0219&6.983&&2.0326&1078\\
Schneider~Electric&0.821&0.0216&6.874&&2.2719&967\\
Soci\'et\'e~G\'en\'erale&0.757&0.0219&7.993&&2.1324&2959\\
Suez&0.719&0.0290&6.774&&2.0518&875\\
TF1&0.920&0.0294&8.865&&2.5094&7079\\
Thales&0.875&0.0272&6.250&&2.3995&518\\
Total&0.594&0.0181&8.930&&1.8116&7553\\
Vinci&0.727&0.0197&7.103&&2.0688&1215\\
Vivendi~Universal&0.888&0.0233&10.491&&2.4294&35983\\
\hline\hline
\end{tabular}
\end{center}
\caption[CAC~40. GMM estimation of the MRW log-normal model]
{\label{table_cacgmm}\small  The first three columns correspond to the estimations of $\ln\sigma$, $\lambda^2$ and $\ln T$ given by the GMM. 
The next two columns 
show the corresponding values of $\sigma$ and $T$. }
\end{small}
\end{table}

One can see that, in most cases, the estimation of $T$ lead to values greater or of the order of $L = 3770$. For those, the low frequency regime is clearly not reached. The observation scale $L$ is not
large enough compared to the integral scale. However, one can consider that the high frequency regime is reached $T \gg  1$. Thus, as shown in section \ref{sec:estgrT}, (i) the estimator of $\sigma$ 
does not converge, (ii) the estimation of $T$ is not reliable and depends essentially on $L$ however, (iii) the estimation of $\lambda^2$ is reliable. The only way to get confidence intervals is to use Monte-Carlo. We
computed $5\%$ confidence interval using $15000$ realizations of the MRW process with $\sigma = 1$ (we normalized the logarithm of the stock prices), $\lambda^2 = 0.02$  and $T = 3770$. We got 
$\widehat{\lambda^2}\in[0.013,0.027]$,
and $\widehat{T}\in[200,250000]$.
This shows that all the results in Table \ref{table_cacgmm} are compatible (at a $5\%$ level) with a single set of parameters : $\lambda^2 = 0.02$ and $T=3770$.

\subsection{Volatility forecasting}
Volatility is a model dependent notion. For instance, for GARCH models \cite{Bol86}, at a given time the conditional volatility (to all the observed past) is a deterministic number whereas for stochastic volatility models as well as  for the MRW model it is a random variable.
In order to compare different models on what is generally referred to as ``volatility forecasting'', one needs to define a common problematic. The problematic we consider here is the  forecasting of absolute returns :
forecasting $|\delta_{s}X(t_{0}+h)|$
knowing all the past data $\{\delta_{\tau}X(t)\}_{t\le t_{0}}$ (with $s\ge \tau = 1$ day).
The parameter $s$ will be referred to as the prediction scale and $h$ as the horizon.

Three methods will be used for volatility forecasting using MRW model. The first one, referred to as MRWLin simply corresponds to solving the linear prediction problem of estimating $\delta_{s}X(t_{0}+h)$ as a linear combination of $\{\delta_{\tau}X(t)\}_{t\le t_{0}}$. For this purpose we need analytical expression of the mean and the auto-covariance of the increments process. 
This is given by Eq. \eqref{eqn:MRWlogn2} of
Theorem \ref{thm:MRWlogn2} along with Eqs \eqref{eq:clogMRM2}, \eqref{eq:clogMRM2b} and \eqref{eq:clogMRM2bis} of Proposition \ref{prop:covmag}.
The second one, referred to as MRWSq, is
based on the exact same equations. It simply corresponds to solving the linear prediction problem of estimating $|\delta_{s}X(t_{0}+h)|^2$ as a linear combination of $\{|\delta_{\tau}X(t)|^2\}_{t\le t_{0}}$. 
The last one, referred to as MRWLog simply corresponds to solving the linear prediction problem of estimating $\ln|\delta_{s}X(t_{0}+h)|$ as a linear combination of $\{\ln|\delta_{\tau}X(t)|\}_{t\le t_{0}}$. For this purpose we need an analytical expression of the mean and the auto-covariance of  the logarithm of the increments process. This is given by Eqs \eqref{eq:covM} and \eqref{eq:mlogMRM} of
Theorem \ref{th:covM} along with Eqs \eqref{eq:clogMRM2}, \eqref{eq:clogMRM2b} and \eqref{eq:clogMRM2bis} of Proposition \ref{prop:covmag}.

We compare the results of the MRW-based forecasting with econometric models that are  standard for volatility forecasting. We use the standard GARCH(1,1) model (with normal innovations) and the t-student GARCH(1,1) model (with t-student innovations) referred to as the tGARCH(1,1) model \cite{AndBolChrDie05}.

For all the MRW-based forecasting, we use the same two parameters $\lambda^2 = 0.02$ and $T = L = 3770$ (in the previous section, we have seen that these values were compatible with the estimations performed on most of the stocks).
 In order to make the GARCH-based forecasting harder to beat, we choose to estimate the GARCH parameters using maximum likelihood estimators  \cite{LeeHan94} {\em separately} on each stock time-series, using the {\em entire} time-series. Thus, we perform in-sample GARCH-based forecasting and out-of-sample MRW-based forecasting. The forecasting errors are computed using both an $L^2$ norm (MSE) and a $L^1$ norm (MAE). Table \ref{tableCres} displays the number of stocks (out of the 29) for which the corresponding MRW-based forecasting beat both the GARCH(1,1) and the tGARCH(1,1) forecasting. This is done independently for each error (MSE or MAE) and for different horizons and scales. We see that for  MAE error all MRW-based forecasting clearly outperform both GARCH(1,1) forecasting at any horizon and any scale. MRWLin performs even better than the two other MRW-based methods.
For MSE error, both MRWLin and MRWSq forecasting outperform both GARCH forecasting with a preference toward MRWSq (which is not surprising since the corresponding linear prediction, by definition, minimizes the mean square root error).

\begin{table}
\begin{small}
\begin{center}
\begin{tabular}{llccc}
\hline
\hline
$s$ & $h$ & MRWLin & MRWSq & MRW~Log\\
\hline
 & & \multicolumn{3}{c}{MAE}\\
\hline
1~day & 1~day & 28 & 26 & 29\\
5~days & 5~days & 28 & 24 & 26\\
10~days & 10~days & 27 & 24 & 19\\
20~days & 20~days & 26 & 23 & 17\\
10~days & 20~days & 28 & 24 & 25\\
20~days & 40~days & 28 & 20 & 23\\
\hline
 & & \multicolumn{3}{c}{MSE}\\
\hline
1~day & 1~day & 13 & 22 & 2\\
5~days & 5~days & 17 & 19 & 2\\
10~days & 10~days & 16 & 19 & 4\\
20~days & 20~days & 21 & 19 & 8\\
10~days & 20~days & 22 & 22 & 8\\
20~days & 40~days & 17 & 16 & 8\\
\hline
\hline
\end{tabular}
\end{center}
\caption{
\label{tableCres}\small Forecasting performance of the three MRW-based methods (MRWLin, MRWSq, MRWLog) depending on the error criterion (either $L^2$ : MSE or $L^1$ : MAE) and on the scale $s$ and horizon $h$ forecasting.
Each entry corresponds to the number of stocks (out of 29) for which the corresponding MRW-based forecasting beat both GARCH(1,1) and tGARCH(1,1)-based forecasting.}
\end{small}
\end{table}

\subsection{Value at Risk forecasting}
Given the log return process $X(n\tau)$ (remember that $\tau = 1$ day  in our case) and the present time $t_{0} = n_{0}\tau$, the conditionnal Value at Risk $\mathrm{VaR}_p$
at confidence level $p$, at scale $s \ge \tau$ and at horizon
$h\ge 0$ is defined by he relation 
\begin{equation}
\mathbb{P}\big[\delta_sX(t_{0}+h)>-\mathrm{VaR}_p(t_{0})\big|X(n\tau),n\leq n_{0}\big]=p.
\end{equation}
It thus corresponds to the maximum loss on a given scale and horizon and at a given confidence level. The highest it is the riskiest the asset is.

The estimation of $VaR_{p}$  in the case $X$ is an MRW process is based on Eq. \eqref{eq:MRWlogn1} of Theorem \ref{thm:MRWlogn1}.
This equation means that the process $\ln|\delta_{\tau}X(n\tau)|$ can be seen (in the first order in $\lambda^2$ as the sum of the logarithm of a white gaussian noise $\epsilon[n]$ and of the renormalized magnitude which is gaussian and independant of $\epsilon[n]$.
Thus, at time $t_{0}$, the conditionnal law of $|\delta_{\tau}X(n\tau)|$ and consequently the associated value at risk can be estimated using an estimation of the conditionnal law 
(i.e., the conditionnal mean and variance) of the renormalized magnitude.  Conditionnal mean and variance estimations can be performed solving, as in the previous section, the linear prediction 
problem of estimating $\ln|\delta_{s}X(t_{0}+h)|$ as a linear combination of $\{\ln|\delta_{\tau}X(t)|\}_{t\le t_{0}}$. The prediction obtained corresponds to the conditionnal mean estimation and the variance of the prediction corresponds to the conditionnal variance estimation.

We use two different tests for testing the estimated conditonnal value at risk. They are both based on the series
\begin{equation}
I_p[n_{0}]=\begin{cases}
1,&\text{if $\delta_\tau X(n_{0}\tau+h)<-\mathrm{VaR}_p$},
\\
0,&\text{if $\delta_\tau X(n_{0}\tau+h)\geq-\mathrm{VaR}_p$},
\end{cases}
\end{equation} 
One can easily show that the process $\{I_{p}[n]\}$ is a Bernoulli process with parameter $p$. 
Thus, the first test, generally called the Kupiec test \cite{Kup95}, is based on the equation
\begin{equation}
\label{hyp:kup}
\mathbb{P}\big[I_p[n]=1\big]=p.
\end{equation}
Thus, this test does not take into account the dynamic of the $\{I_{p}[n]\}$.
The Christoffersen test \cite{Chr98} does. It is based on the fact that 
\begin{equation}
\label{hyp:chr}
\mathbb{P}\big[I_p[n]=1\big|I_p[n-1]=0\big]=\mathbb{P}\big[I_p[n]=1\big|I_p[n-1]=1\big]=p.
\end{equation}

\begin{table}
\begin{small}
\begin{center}
\begin{tabular}{lccc}
\hline
\hline
  & nGARCH & tGARCH & MRW\\
\hline
$p$  & \multicolumn{3}{c}{$s$ = 1~day,\quad $h$ = 1~day}\\
\hline
0.5\% & 11, 15 & 2, 4 & {\bf 24}, {\bf 23}\\
1\% & 21, 21 & 0, 2 & {\bf 21}, 20\\
5\% & 16, 13 & 0, 0 & {\bf 22}, {\bf 13}\\
10\% & 3, 5 & 0, 0 & {\bf 24}, {\bf 20}\\
20\% & 2, 2 & 0, 0 & {\bf 25}, {\bf 24}\\
\hline
 $p$& \multicolumn{3}{c}{$s$ = 1~day,\quad $h$ = 6~days}\\
\hline
0.5\% & 7, 10 & 2, 7 & {\bf 26}, {\bf 20}\\
1\% & 22, 15 & 1, 1 & {\bf 22}, {\bf 16}\\
5\% & 11, 7 & 0, 0 & {\bf 20}, {\bf 10}\\
10\% & 2, 3 & 0, 0 & {\bf 21}, {\bf 12}\\
20\% & 0, 1 & 0, 0 & {\bf 25}, {\bf 18}\\
\hline
 $p$& \multicolumn{3}{c}{$s$ = 1~day,\quad $h$ = 11~days}\\
\hline
0.5\% & 9, 10 & 5, 9 & {\bf 26}, {\bf 19}\\
1\% & 22, 15 & 1, 1 & {\bf 24}, 14\\
5\% & 11, 6 & 0, 0 & {\bf 20}, {\bf 8}\\
10\% & 2, 3 & 0, 0 & {\bf 22}, {\bf 9}\\
20\% & 0, 0 & 0, 0 & {\bf 26}, {\bf 19}\\
\hline
 $p$& \multicolumn{3}{c}{$s$ = 5~days,\quad $h$ = 5~days}\\
\hline
0.5\% & 22, 21 & 22, 22 & {\bf 23}, 21\\
1\% & 23, 24 & 14, 19 & {\bf 26}, {\bf 24}\\
5\% & 23, 24 & 4, 4 & 21, {\bf 24}\\
10\% & 19, 18 & 4, 4 & {\bf 23}, {\bf 24}\\
20\% & 14, 17 & 2, 7 & {\bf 21}, {\bf 22}\\
\hline
 $p$& \multicolumn{3}{c}{$s$ = 5~days,\quad $h$ = 10~days}\\
\hline
0.5\% & 25, 20 & 23, 18 & {\bf 26}, 18\\
1\% & 26, 25 & 14, 17 & 24, 23\\
5\% & 22, 19 & 3, 4 & {\bf 23}, {\bf 24}\\
10\% & 15, 18 & 1, 4 & {\bf 24}, {\bf 24}\\
20\% & 10, 14 & 0, 4 & {\bf 21}, {\bf 22}\\
\hline
\hline
\end{tabular}
\end{center}
\caption[CAC~40. R\'esum\'e de la pr\'ediction de VaR]
{\label{tableCres2}
\small Number of stocks (out of 29) that are accepted by the Kupiec test (left number)  and the Christoffersen test (right number) for a confidence level of $95\%$.
Bold face numbers correspond to the case where the MRW-based estimation passes the test  more times than both GARCH and tGARCH based estimation. 
}
\end{small}
\end{table}
The results of these tests are illustrated in Table \ref{tableCres2} for both MRW-based estimation (using the estimation of the renormalized magnitude conditionnal law as described above) and GARCH-based estimation \cite{}.
As in the previous section, the MRW parameters $\lambda^2$ and $T$ are fixed once for all ($\lambda^2=0.02$, $T=3770$) whereas, the GARCH (and tGARCH) parameters are estimated using a maximum likelihood estimation 
independantly for {\em each} stock on the {\em entire} time-series. Each entry of Table \ref{tableCres2} is composed of two integers separated by a comma. The left number is the number of stocks (out of 29) which passed the Kupiec test, the right number to the number of stocks that passed the Christoffersen test (both tests are performed using a confidence level of $95\%$). Bold face numbers correspond to the case where the MRW-based estimation passes the test  more times than both GARCH and tGARCH based estimation. 

The superiority of the MRW-based estimation over the GARCH or tGARCH-based estimation appears very clearly at all scales and horizons and at all level $p$.

\section{Conclusions and prospects}
\label{cp}
In this paper, we have reviewed the main properties of log-normal continuous cascades and developed an approximation
framework in the limit of small intermittency. We have shown that, within this approximation, the law of the the process increments, at each time scale, 
can be expressed under simple analytic forms so that
the process aggregation properties are easy to control and deal with.
As far as parameter estimation problems are concerned, we have pointed out that one has to distinguish ``low'' and ``high'' frequency
asymptotic regimes according to which the properties of the samples are somehow different. In the ``low'' frequency regime, one considers samples of abritrary increasing sizes at fixed sampling frequency. 
In that case, we have shown that the model parameters can be estimated 
with a GMM method mainly relying upon empirical covariance function of log-increments of the processes. The high frequency regime corresponds to a situation when the process is observed 
over a finite lenght and sampled at increasing rate. 
This case is not equivalent to the former one and only the intermittency p
aramter $\lambda^2$ can be faithfully estimated.
Indeed, because of the self-similarity of the process, the integral scale is no longer a parameter 
and can arbitrarily chosen to be the overall sample lenght while the estimator 
of $\sigma^2$ converges towards a random value.

Our approach has been applied to financial time series for which it is well known that
log-normal MRW provides a particularily parcimonious model that 
allows one to reproduce most
of well documented stylized facts. At $5 \%$ confidence level, our estimates show that
all the analyzed stock return series are multifractal but with a small intermittency coefficient
$\lambda^2 \simeq 0.02$. The low intermittency $\lambda^2 \ll 1$ approximation is thus likely to be sound. 
Moreover, our estimates of $T$ values 
suggest that the integral scale magnitude order is greater than one or several years. 
The ability of log-normal MRW to model volatility dynamics has been illustrated by its
perfomances in conditional Value at Risk forecasting.
From a practical point of view, the main interest of MRW-like models 
is that they capture the ''heteroskedastic'' nature of 
return fluctuations, by preserving, in some sense, 
the nice stability properties accross time scales of the Brownian motion. 

Time series analysis involving multifractal processes is still in its
infancy. In forthcoming studies, we will extend the approach presented here 
to continuous cascades with arbitrary log-infinitely divisible laws.
In particular the prospect to define a semi-parametric test for the multifractal nature
of a time series is very appealing.
The two asymptotic regimes discussed in this paper, also challenge many 
interesting issues: in some recent work, we have shown that they
can be described within the general framework of ``mixed asymptotics''.
In this regime, the overall sample lenght $L$ increases while the sampling scale 
$\tau \rightarrow 0$\cite{MuzBacKoz06, BacKozMuz06,MuzBacBaiPog08, BacGlotHofMuz08}.
A priori many statistics related to multifractal processes and notably the 
extreme value statisitics explicitely depend which ``mixed asymptotics'' we are, i.e., 
on the relative velocity according to which $L \rightarrow \infty$ and 
$\tau \rightarrow 0$ \cite{MuzBacKoz06}.

\appendices

\section{Existence of the renormalized magnitude $\Omega(t)$}
\label{app:renmag}

\begin{Lm}[Convergence of the finite dimensional laws of $\Omega_{l}(t)$]
\label{lm:conloifin}
Let $t_1,\ldots,t_n$, $n$ be real numbers, then the Gaussian vector 
$\big(\Omega_l(t_1),\ldots,\Omega_l(t_n)\big)$ converges, when $l$ goes to 
zero, toward the centered Gaussian vector $\big(\Omega(t_1),\ldots,\Omega(t_n)\big)$, which does not depend on 
$\lambda$ and which covariance matrix reads:
\begin{equation}
\label{eq:vcvOm}
(\Sigma)_{jk}={\mathbb C}{\mathrm ov}\big[\Omega(t_j),\Omega(t_k)\big]
=\frac 1 {\lambda^2} \int\limits_0^{t_j}du\int\limits_0^{t_k}dv\rho(u-v),
\end{equation}
where the function $\rho(t)$ is defined by
\begin{equation}
\label{eq:rho}
\rho(t)=\begin{cases}
\lambda^2\ln\Big(\frac{T}{|t|}\Big),&\text{if $|t|< T$},
\\
0,&\text{if $T\leq |t|$}.
\end{cases}
\end{equation}
\end{Lm}

\begin{IEEEproof}
The function $\rho_l(t)$ being defined (cf. section \ref{2}) 
as the correlation function of the process $\omega_{l,T}(t)$, the vector 
$\big(\Omega_l(t_1),\ldots,\Omega_l(t_n)\big)$ is a centered Gaussian vector
which covariance matrix $\Sigma_l$ is
\begin{equation}
(\Sigma_l)_{jk}={\mathbb C}{\mathrm ov}\big[\Omega_l(t_j),\Omega_l(t_k)\big]
= \frac{1}{\lambda^2}\int\limits_0^{t_j}du\int\limits_0^{t_k}dv\rho_l(|u-v|),
\end{equation}

This matrix converges toward the matrix $\Sigma$ (which all coefficients
are finite) when $l \to 0$. It thus suffices to show
that $\Sigma$ is semi-defined positive. 
It can be shown \cite{Koz06} that the function
$\rho(t)$ can be written as
\begin{equation}
\rho(t)=\int\limits_{-\infty}^{\infty}{\tilde \rho}(t-s){\tilde \rho}(s)ds.
\end{equation}
If one defines the vector $V(s)$ as
\begin{equation}
V(s)=\Big(\int\limits_0^{t_1}{\tilde\rho}(u_1-s)du_1,\ldots,
\int\limits_0^{t_n}{\tilde \rho}(u_n-s)du_n\Big)^{T}
\end{equation}
then the matrix $\Sigma$ can be written as
\begin{equation}
\Sigma=\int\limits_{-\infty}^{\infty}V(s)V^T(s)ds,
\end{equation}
thanks to the identity ${\tilde \rho}(s)={\tilde \rho}(-s)$. 
Consequently, $\Sigma$ semi-defined positive.
\end{IEEEproof}

Because the variables $\Omega_{l}(t)$ and $\Omega(t)$ are Gaussian, in
order to show the tighness of the sequence 
$\{e^{2\Omega_{l}(t)}\}_t$, it is sufficient to show the following 
proposition:

\begin{Lm}[Tightness]
\label{lm:tightness}
It exists $\epsilon \in [0,1[$ such that
\begin{equation}
\label{eq:tens}
\EE{\big(e^{2\Omega(t)}-e^{2\Omega(s)}\big)^2}=o\left((t-s)^{2-\epsilon}\right),
~\text{$\forall t,s$}.
\end{equation}
\end{Lm}

\begin{IEEEproof}
A direct computation leads to the following equation:
\begin{multline}
\label{eq:tens2}
\EE{\big(e^{2\Omega(t)}-e^{2\Omega(s)}\big)^2}
=e^{8{\mathbb V}{\mathrm ar}[\Omega(t)]}+e^{8{\mathbb V}{\mathrm ar}[\Omega(s)]}
-2e^{2{\mathbb V}{\mathrm ar}[\Omega(t)+\Omega(s)]}
\\
=\left(e^{4{\mathbb V}{\mathrm ar}[\Omega(t)]}-e^{4{\mathbb V}{\mathrm ar}[\Omega(s)]}\right)^2
+2e^{4{\mathbb V}{\mathrm ar}[\Omega(t)]+4{\mathbb V}{\mathrm ar}[\Omega(t)]}
\left(1-e^{-2{\mathbb V}{\mathrm ar}[\Omega(|t-s|)]}\right).
\end{multline}

The first term of \eqref{eq:tens2} can be estimated as
\begin{eqnarray*}
\left(e^{4{\mathbb V}{\mathrm ar}[\Omega(t)]}-e^{4{\mathbb V}{\mathrm ar}[\Omega(s)]}\right)^2 & = & 64t^2e^{8t^2\ln\big(\frac{Te^{3/2}}{t}\big)}(t-s)^2+o((t-s)^2) \\
& = & o\left((t-s)^{2-\epsilon}\right).
\end{eqnarray*}
and the second term as 
\begin{eqnarray*}
2e^{4{\mathbb V}{\mathrm ar}[\Omega(t)]+4{\mathbb V}{\mathrm ar}[\Omega(t)]}
\left(1-e^{-2{\mathbb V}{\mathrm ar}[\Omega(|t-s|)]}\right)
 & = & 2e^{8t^2\ln\big(\frac{Te^{3/2}}{t}\big)}\ln\bigg(\frac{Te^{3/2}}{|t-s|}\bigg)(t-s)^2 
+o((t-s)^2) \\
& = & o\left((t-s)^{2-\epsilon}\right).
\end{eqnarray*}
\end{IEEEproof}

\begin{Lm}
\label{lm:magmoments}
If  $I_1,\ldots,I_n$, be $n$ arbitrary intervals then
\begin{equation}
\EE{\frac{\Omega(I_1)}
{|I_1|},\ldots,\frac{\Omega(I_n)}{ |I_n|}} =  K(I_{1},\ldots,I_{n}),
\end{equation}
where $K(I_{1},\ldots,I_{n})$ reads 
\begin{equation}
\label{eq:Kn}
K(I_{1},\ldots,I_{n})=\begin{cases}
\!\!\!\sum\limits_{P({\mathcal I}_n)}\!\Big[
\prod\limits_{k=1}^{n/2}\int\limits_{I_{i_k}}\!\frac{du_{i_k}}{|I_{i_k}|}\int\limits_{I_{j_k}}\!%
\frac{du_{j_k}}{|I_{j_k}|}\rho(u_{i_k}-u_{j_k}),&\text{\!\!\!\!if $n$ is even},
\\
0,&\text{\!\!\!\!otherwise},
\end{cases}
\end{equation}
where $\rho(t)$ is defined by Eq \eqref{eq:rho} and
$P({\mathcal I}_n)$ is the set of all non ordinated partitions of two elements 
of ${\mathcal I}_n$. An element of  $P({\mathcal I}_n)$ can thus be written as 
$\{(i_k,j_k)\}_{k=1,\ldots,n/2}$.
\end{Lm}
\begin{IEEEproof}
This result can be obtained from a direct computation relying upon Wick's Theorem\cite{Wic50} and Lemma \ref{lm:conloifin} of Appendix~\ref{app:renmag} 
\end{IEEEproof}

\section{Taylor expansion of the moments of the measure}
\label{appsi}
The following proposition links the centered generalized moments of $M$ with those of $\Omega$. 
It will be used to prove limit theorems of sections \ref{ssec:law} and \ref{ssec:amlm}.
\begin{Pp}
\label{thm:momcgen}
Let $n$ some positive integer. The generalized centered moment of the log-normal MRM
measure of the intervals $I_1,\ldots,I_n$ admits the following Taylor series expansion 
when $\lambda^2 \rightarrow 0$:
\begin{equation}
\label{eq:momcgen2}
\EE{\bigg(\frac{M(I_1)}{|I_1|}-1\bigg)\cdots
\bigg(\frac{M(I_n)}{|I_n|}-1\bigg)}=2^n \lambda^n \EE{\frac{\Omega(I_1)}
{|I_1|},\ldots,\frac{\Omega(I_n)}{ |I_n|}} + o(\lambda^n), 
\end{equation}
where $\Omega$ is the renormalized magnitude defined in Section \ref{ssection:renmag}.
\end{Pp}
\begin{IEEEproof}
Let us note that the right handside of Eq. \eqref{eq:momcgen2} is given by Lemma \ref{lm:magmoments} of Appendix \ref{app:renmag}. We are going to prove that the left handside of Eq. \eqref{eq:momcgen2} is equal to the same expression.

Let us first indroduce the following random variables:
\begin{equation}
M_j=\frac{M(I_j)}{|I_j|},~\text{pour $j=1,\ldots,n$}.
\end{equation}

The centered generalized moment corresponding to intervals
$I_1,\ldots,I_n$ can be written as a linear combination of 
generalized moments:
\begin{equation}
\EE{(M_1-1)\cdots(M_n-1)}
=\sum_{m=0}^n(-1)^{n-m}\sum_{{\mathcal I}_m\subset{\mathcal I}_n}
\EE{M_{i_1}\cdots M_{i_m}}.
\end{equation}

From integral represention \eqref{eq:momgen}, we have:

\begin{equation}
\label{eq:expr}
\EE{(M_1-1)\cdots(M_n-1)}
=\sum_{m=0}^n(-1)^{n-m}\sum_{{\mathcal I}_m\subset{\mathcal I}_n}
\int\limits_{I_{i_1}}\frac{du_{i_1}}{|I_{i_1}|}\ldots
\int\limits_{I_{i_m}}\frac{du_{i_m}}{|I_{i_m}|}e^{\lambda^2{\mathcal S}({\mathcal I}_m)},
\end{equation}
where ${\mathcal S}({\mathcal I}_m)$ is the following symmetric sum;
\begin{equation}
{\mathcal S}({\mathcal I}_m)=\sum_{\substack{i_j,i_k\in{\mathcal I}_m\\i_j<i_k}}X_{i_ji_k},
\end{equation}
with
\begin{equation}
X_{i_ji_k}=4\ln\left(\frac{T}{|u_{i_j}-u_{i_k}|}\right)
{\mathbf 1}_{\{|u_{i_j}-u_{i_k}|\leq T\}}.
\end{equation}

It is possible to integrate previous expression as respect $u_1,\ldots,u_n$: 
\begin{equation}
\int\limits_{I_{i_1}}\frac{du_{i_1}}{|I_{i_1}|}\ldots
\int\limits_{I_{i_m}}\frac{du_{i_m}}{|I_{i_m}|}e^{\lambda^2{\mathcal S}({\mathcal I}_m)}
=\int\limits_{I_1}\frac{du_1}{|I_1|}\ldots\int\limits_{I_n}\frac{du_n}{|I_n|}
e^{\lambda^2{\mathcal S}({\mathcal I}_m)}.
\end{equation}
and changing the order of integration leads to:
\begin{equation}
\EE{(M_1-1)\cdots(M_n-1)}=\int\limits_{I_1}\frac{du_1}{|I_1|}\ldots
\int\limits_{I_n}\frac{du_n}{|I_n|}\sum_{m=0}^n(-1)^{n-m}\sum_{{\mathcal I}_m\subset{\mathcal I}_n}
e^{\lambda^2{\mathcal S}({\mathcal I}_m)}.
\end{equation}

In \cite{Koz06} it is shown that the generalized moments as functions 
of $\lambda^2$ belong to $C^{\frac{n}{2}}(\RR)$ the class
of $\frac{n}{2}$ times continuously differentiable functions. 
The  $k$-th derivative of the generalized centered moment as respect to 
$\lambda^2$ in $\lambda^2 = 0$ reads:
\begin{equation}
\label{eq:expr2}
\Big(\frac{\partial}{\partial\lambda^2}\Big)^{k}
\EE{(M_1-1)\cdots(M_n-1)}\Big|_{\lambda^2=0}
=\int\limits_{I_1}\frac{du_1}{|I_1|}\ldots\int\limits_{I_n}\frac{du_n}{|I_n|}
\sum_{m=0}^n(-1)^{n-m}\sum_{{\mathcal I}_m\subset{\mathcal I}_n}\big({\mathcal S}({\mathcal I}_m)\big)^k
\end{equation}

Let us consider some arbitrary integer $j$. One can regroup the terms under the integral in \eqref{eq:expr2}
\begin{equation}
\label{eq:expr3}
\sum_{m=0}^n(-1)^{n-m}\!\!\sum_{{\mathcal I}_m\subset{\mathcal I}_n}\big({\mathcal S}({\mathcal I}_m)\big)^k
=\sum_{m=0}^{n-1}(-1)^{n-m}\!\!\!\!\!\sum_{{\mathcal I}_m\subset{\mathcal I}_n\setminus\{j\}}\!
\big({\mathcal S}({\mathcal I}_m)\big)^k-\big({\mathcal S}({\mathcal I}_m\cup\{j\})\big)^k
\end{equation}

By noticing that the symmetric sum ${\mathcal S}({\mathcal I}_m\cup\{j\})$ can be rewritten as
\begin{equation}
{\mathcal S}({\mathcal I}_m\cup\{j\})={\mathcal S}({\mathcal I}_m)+{\mathcal C}(j,{\mathcal I}_m),
\end{equation}
where ${\mathcal C}(j,{\mathcal I}_m)$ is defined as
\begin{equation}
{\mathcal C}(j,{\mathcal I}_m)=\sum_{i_k\in I_m}X_{j i_k},
\end{equation}
we can see that
\begin{equation}
\big({\mathcal S}({\mathcal I}_m)\big)^k-\big({\mathcal S}({\mathcal I}_m\cup\{j\})\big)^k
=-{\mathcal C}(j,{\mathcal I}_m)\sum_{i=1}^{k}\frac{k!}{i!(k-i)!}
\big({\mathcal C}(j,{\mathcal I}_m)\big)^{i-1}\big({\mathcal S}({\mathcal I}_m)\big)^{k-i}.
\end{equation}

This last relationship means that each term of the sum \eqref{eq:expr3} contains at least
one factor like $X_{j i_m}$. Because $j$ is arbitrarily fixed, the sets of factor indices of each
term of the sum \eqref{eq:expr3} must contain
all indices $1,\ldots,n$. Therefore, if the derivative order $k$ is smaller than
$n/2$, the sum \eqref{eq:expr3} must vanish. It results that the first non trivial order 
in the Taylor series in power of $\lambda^2$ of the centered generalized moment is at least 
$\frac{n+1}{2}$ if $n$ est odd, and $\frac{n}{2}$ if $n$ is even. 
In this latter case, the first order terms in the Taylor series are proportionnal to 
$\lambda^{n}X_{i_1j_1}\cdots X_{i_{n/2}j_{n/2}}$, where the set of indices 
$\{i_1,j_1,\ldots,i_{n/2},j_{n/2}\}$ contains all the values $1,\ldots,n$. Such terms
can only come from the expansion of 
$e^{\lambda^2{\mathcal S}({\mathcal I}_n)}$, which leads to the expected result.
\end{IEEEproof}

The following Lemma will be used to prove the limit theorem of Section \ref{ssec:amm}.
\begin{Lm}
\label{lm:fl} 
For some arbitrary intervals $I_1$ \ldots, $I_n$, the generalized
moments of a log-normal MRM measure has the following integral representation:
\begin{equation}
\label{eq:momgen}
\EE{\frac{M(I_1)}{|I_1|}\cdots \frac{M(I_n)}{|I_n|}}
=\int\limits_{I_1}\frac{du_1}{|I_1|}\ldots\int\limits_{I_n}\frac{du_n}{|I_n|}
f_{\lambda^2}(u_1,\ldots,u_n),
\end{equation}
where the function $f_{\lambda^2}(u_1,\ldots,u_n)$ is
\begin{equation}
\label{eq:deff}
f_{\lambda^2}(u_1,\ldots,u_n)=\exp\bigg(4\lambda^2\sum_{1\leq i<j\leq n}
\rho(u_{i}u_{j})\bigg),
\end{equation}
where $\rho(t)$ is defined by Eq \eqref{eq:rho}.
\end{Lm}
\begin{IEEEproof}
This directly results from the fact that $\lim_{l\rightarrow 0^+} \EE{M_{l,T}[0,t]} = \EE{M_{T}[0,t]} = t$ \cite{BacMuz03}  and some simple algebra (see \cite{Koz06}).
\end{IEEEproof}
For the sake of simplicity let us introduce the following set of indices:
\begin{equation}
{\mathcal I}_m=\{i_1,\ldots,i_m\}\subset{\mathcal I}_n=\{1,\ldots,n\}.
\end{equation}

\section{Taylor expansion of the moments of the logarithm of the measure}
\label{app:log}
In this section we establish some results useful to prove limit theorems of Sections \ref{ssec:amlm} and \ref{ssec:law}.
According to Proposition \ref{thm:mmomM}, all the moments of the MRM of negative orders are finite, consequently the moments of the logarithm of the MRM are also finite.

Along the same line as in previous appendix, one can write an expansion for the 
generalized moment of the logarithm of a log-normal MRM measure.

\begin{Pp}
\label{prop:momloggen}
Let $n$ be a positive integer. The generalized centered moment of the logarithm of the MRM measure 
of intervals $I_1,\ldots,I_n$ admits the following Taylor series expansion around $\lambda^2 = 0$: 
\begin{equation}
\label{eq:momloggen}
\EE{\ln\left(\frac{M(I_1)}{|I_1|}\right)\cdots
\ln\left(\frac{M(I_n)}{|I_n|}\right)} = \EE{\bigg(\frac{M(I_1)}{|I_1|}-1\bigg)\cdots
\bigg(\frac{M(I_n)}{|I_n|}-1\bigg)} + o(\lambda^n),
\end{equation}
where $K(I_{1},\ldots,I_{n})$ are defined Eq. \eqref{eq:Kn}. 
Let us recall that $K(I_{1},\ldots,I_{n}) = 0$ if $n$ is odd.
\end{Pp}
The proof of this Proposition is postponed to the end of this section, it is based on the following Lemma \ref{lem:log1}, \ref{lem:log2} and \ref{lem:log3} :
\begin{Lm}
\label{lem:log1}
Let $0<\epsilon<1$ and $m$ be a positive integer.
One has the following inequality:
\begin{equation}
\label{eq:bornprob}
{\mathbb P}\big[M\not\in B_{\epsilon}\big]\leq o(\lambda^{2m}),
\end{equation}
where the compact subset of ${\mathbb R}^n$, $B_{\epsilon}$ is defined by:
\begin{equation}
\label{def:beps}
B_{\epsilon}=\Big\{x=(x_1,\ldots,x_n)\in{\mathbb R}^n;
\max_{1\leq k\leq n}(|x_k-1|)\leq\epsilon\Big\}.
\end{equation}
and $o(\lambda^{2m})$ depends on de $\epsilon$.
\end{Lm}

\begin{IEEEproof}
Since $\lambda^2 \rightarrow 0$, then, without loss of generality, one can assume that
the order $2m+2$ centered moment of the log-normal MRM measure exists 
(see Theorem \ref{thm:momM}). Thanks to Bienaym\'e-Tchebychev inequality, one has:
\begin{equation}
{\mathbb P}\big[M\not\in B_{\epsilon}\big]
\leq\sum_{j=1}^n{\mathbb P}\big[\left|M_j-1\right|>\epsilon\big]
\leq\sum_{j=1}^n\frac{1}{\epsilon^{2m+2}}{\mathbb E}\big[(M_j-1)^{2m+2}\big]=o(\lambda^{2m}).
\end{equation}
\end{IEEEproof}

\begin{Lm}
\label{lem:log2}
Let $0<\epsilon<1$. For all continuous function $f(M)$ 
over the compact set $B_{\epsilon}$ defined in \eqref{def:beps}, 
the following inequality holds:
\begin{equation}
\EE{f(M)}-\EE{f(M)\big|M\in B_{\epsilon}}
\leq\sqrt{\EE{f(M)^2}}\sqrt{{\mathbb P}\big[M\not\in B_{\epsilon}\big]}
+\Big(\sup_{M\in B_{\epsilon}}|f(M)|\Big){\mathbb P}\big[M\not\in B_{\epsilon}\big].
\end{equation}
\end{Lm}

\begin{IEEEproof}
From the law of total probabilities, it follows that
\begin{eqnarray*}
\EE{f(M)}-\EE{f(M)\big|M\in B_{\epsilon}}
& = & \EE{f(M)\big|M\not\in B_{\epsilon}}{\mathbb P}[M\not\in B_{\epsilon}]
+\EE{f(M)\big|M\in B_{\epsilon}}\big({\mathbb P}[M\in B_{\epsilon}]-1\big)
\\
 & = & \EE{f(M)\big|M\not\in B_{\epsilon}}{\mathbb P}\big[M\not\in B_{\epsilon}\big]
-\EE{f(M)\big|M\in B_{\epsilon}}{\mathbb P}\big[M\not\in B_{\epsilon}\big],
\end{eqnarray*}
where the first term can be bounded using Cauchy-Schwartz inequality
\begin{equation}
\EE{f(M)\big|M\not\in B_{\epsilon}}{\mathbb P}\big[M\not\in B_{\epsilon}\big]
= \EE{f(M)\mathbf{1}_{\{M\not\in B_{\epsilon}\}}}
\leq\sqrt{\EE{f(M)^2}}\sqrt{{\mathbb P}\big[M\not\in B_{\epsilon}\big]},
\end{equation}
whereas the second term is bounded by the supremum over the compact set $B_{\epsilon}$
\begin{equation}
\EE{f(M)\big|M\in B_{\epsilon}}{\mathbb P}\big[M\not\in B_{\epsilon}\big]
\leq\Big(\sup_{M\in B_{\epsilon}}|f(M)|\Big){\mathbb P}\big[M\not\in B_{\epsilon}\big].
\end{equation}
\end{IEEEproof}

\begin{Lm}
\label{lem:log3}
Let $n$ be a positive integer. In the compact set $B_{\epsilon}$, with $\epsilon < 1$, 
one has the following identity of Taylor series expansions up to order $n$ around $\lambda^2 = 0$:
\begin{equation}
\label{eq:tay}
\EE{\prod_{j=1}^{n}\ln(M_j)\Big|M\in B_{\epsilon}}
= \EE{\prod_{j=1}^{n}(M_j-1)\Big|M\in B_{\epsilon}}+o(\lambda^{n})
\end{equation}
\end{Lm}

\begin{IEEEproof}
Thanks to the identity $\ln(x)=x-1-(x-1-\ln(x))$, one has 
\begin{equation}
\prod_{j=1}^{n}\ln(M_j)-\prod_{j=1}^{n}(M_j-1)
=\prod_{j=1}^{n}\big(M_j-1-(M_j-1-\ln(M_j))\big)-\prod_{j=1}^{n}(M_j-1).
\end{equation}
The expansion of the first product of r.h.s. leads to a linear combination of terms such as
$\prod_{j=1}^{k}(M_{i_j}-1-\ln(M_{i_j}))\prod_{j=k+1}^{n}(M_{i_j}-1)$ 
with $1\leq k\leq n$ which conditional expectation can be bounded using Cauchy-Schwartz inequality and
the following inequality established in \cite{Koz06}:
\begin{equation}
\label{eq:ineg5}
\frac{\epsilon-\ln(1+\epsilon)}{\epsilon^2}x^2\leq x-\ln(1+x)
\leq -\frac{\epsilon+\ln(1-\epsilon)}{\epsilon^2}x^2,
~\text{for all $0<\epsilon<1$ and $x\in[-\epsilon,\epsilon]$}.
\end{equation}. 
It results that

\begin{multline}
\EE{\prod_{j=1}^{k}(M_{i_j}-1-\ln(M_{i_j}))
\prod_{j=k+1}^{n}(M_{i_j}-1)\Big|M\in B_{\epsilon}}
\\
\leq\EE{\prod_{j=1}^{k}(M_{i_j}-1-\ln(M_{i_j}))^2\Big|M\in B_{\epsilon}}^{1/2}
\EE{\prod_{j=k+1}^{n}(M_{i_j}-1)^2\Big|M\in B_{\epsilon}}^{1/2}
\\
\leq C_{\epsilon}\EE{\prod_{j=1}^{k}(M_{i_j}-1)^4\Big|M\in B_{\epsilon}}^{1/2}
\EE{\prod_{j=k+1}^{n}(M_{i_j}-1)^2\Big|M\in B_{\epsilon}}^{1/2}.
\end{multline}
Using \eqref{eq:maj2}, with $n=2$ and $n=4$, it is possible to 
remove the condition $M \in B_{\epsilon}$
\begin{multline}
\EE{\prod_{j=1}^{k}(M_{i_j}-1-\ln(M_{i_j}))
\prod_{j=k+1}^{n}(M_{i_j}-1)\Big|M\in B_{\epsilon}}
\\
=\big({\mathcal O}(\lambda^{4k})+o(\lambda^{4n})\big)^{1/2}\big({\mathcal O}(\lambda^{2n-2k})+o(\lambda^{2n})\big)^{1/2}
={\mathcal O}(\lambda^{n+k})=o(\lambda^n).
\end{multline}
\end{IEEEproof}

We are now ready to give the proof of Proposition \ref{prop:momloggen}:

\begin{IEEEproof}
[Proof of Proposition  \ref{prop:momloggen}]
For simplicity purpose let us introduce the random variables:
\begin{equation}
M_j=\frac{M(I_j)}{|I_j|},~\text{for $j=1,\ldots,n$}.
\end{equation}

The major difficulty of the proof relies in the fact that the Taylor series 
of $\ln(1+x)$ around $x=0$ converges in the interval $(-1,1)$.
Let $0<\epsilon<1$ and let us consider thevector
$M=(M_1,\ldots,M_n)$ and the compact set $B_{\epsilon}$ defined in \eqref{def:beps}.
Let us also define the two remaining parts $R^{\text{ln}}(\lambda)$ and $R^{\text{c}}(\lambda)$ as: 
\begin{align}
\label{eq:cond1}
&R^{\text{ln}}(\lambda)=\EE{\ln(M_1)\cdots\ln(M_n)}
-\EE{\ln(M_1)\cdots\ln(M_n)\big|M\in B_{\epsilon}},
\\
\label{eq:cond2}
&R^{\text{c}}(\lambda)=\EE{(M_1-1)\cdots(M_n-1)}
-\EE{(M_1-1)\cdots(M_n-1)\big|M\in B_{\epsilon}}.
\end{align}

By applying Lemma \ref{lem:log2} to $R^{\text{ln}}(\lambda)$ 
and $R^{\text{c}}(\lambda)$, it follows
\begin{align}
\label{eq:estrest1}
&\big|R^{\text{ln}}(\lambda)\big|\leq\sqrt{\EE{\ln(M_1)^2\cdots\ln(M_n)^2}}
\sqrt{{\mathbb P}\big[M\not\in B_{\epsilon}\big]}+|\ln(1-\epsilon)|^n
{\mathbb P}\big[M\not\in B_{\epsilon}\big],
\\
\label{eq:estrest2}
&\big|R^{\text{c}}(\lambda)\big|\leq\sqrt{\EE{(M_1-1)^2\cdots(M_n-1)^2}}
\sqrt{{\mathbb P}\big[M\not\in B_{\epsilon}\big]}+\epsilon^n{\mathbb P}\big[M\not\in B_{\epsilon}\big].
\end{align}

Using the analytical expression of the $q$-order moment of $M$ (cf Proposition \ref{prop:escaleinv}),
one shows that the expectation term  in 
\eqref{eq:estrest1} is uniformely bounded in $\lambda^2$, for $\lambda^2$ smaller than a (small enough) given $\lambda^2_0$.  In \cite{Koz06}, it is  shown that the  the expectation term  in 
\eqref{eq:estrest2} is also uniformely bounded for $\lambda^2<\lambda_{0}^2$.
Using Lemma \ref{lem:log1} with $m=n$, one gets the following inequalities:
\begin{align}
\label{eq:maj1}
&\big|R^{\text{ln}}(\lambda)\big|\leq C\lambda^{n+1},
\\
\label{eq:maj2}
&\big|R^{\text{c}}(\lambda)\big|\leq C\lambda^{n+1},
\end{align}
where $C$ depends on $\epsilon$.

According to \ref{lem:log3}, from definitions \eqref{eq:cond1} and 
\eqref{eq:cond2} and the bounds \eqref{eq:maj1} and \eqref{eq:maj2}, 
we have:
\begin{equation}
\EE{\ln(M_1)\cdots\ln(M_n)}
-\EE{(M_1-1)\cdots(M_n-1)}=o(\lambda^n).
\end{equation}
\end{IEEEproof}

\section{Proof of the consistency of the estimator \eqref{eq:estcint} in the high frequency regime}
\label{appconsist}
In this section we provide of ``simplified'' proof of the consistency of the estimator defined in Eq. \eqref{eq:estcint}.
More presisely, we prove the second part of theorem \ref{thm:varestl2MRM} claiming that
if $n,n'$ are different integers such that $0<n\tau<T$ and $0<n'\tau<T$, then  
in the high frequency asymptotic regime $\tau \rightarrow 0$,
the estimator defined by \eqref{eq:estcint} is consistent and its variance decreases as
\begin{equation}
\label{eq:majvarMRMbis}
{\mathbb V}{\mathrm ar}\big[\widehat{\lambda^2}\big]= {\mathcal O}\bigg(\frac{\ln(N)}{N}\bigg),
\end{equation}
(with $N = L/ \tau$).

\begin{IEEEproof}
In the following we set 
\begin{equation}
Z_{l,\tau}[n]=\ln |\delta_\tau \tilde X_{l,T}(n\tau)|,
\end{equation}
where $ \tilde X_{l,T}$ is the linear-wise process defined in Section \ref{montecarlo}. It is defined by Eq. \eqref{xtilde} 
which uses the measure $\tilde M_{l,T}$ defined by \eqref{mtilde}.
Let us recall that $\tilde X_{l,\tau}(t)$ (resp. $\tilde M_{l,T}(dt)$) converges 
towards $X_T(t)$ (resp. $M_T(dt)$)  when $l\rightarrow 0$  \cite{BacMuz03}.
With no loss of generality, we choose $l$ such that $l/\tau$ is an integer.
Notice that according to Eq. \eqref{xtilde} one has:
\begin{equation}
  \{ \delta_\tau \tilde X_{l,T}(n \tau) \}_n \operatornamewithlimits{=}_{Law} \{ \epsilon[n] \sqrt{\delta_\tau \tilde M_{l,T}(n \tau)} \}_n ,
\end{equation}
where $\epsilon[n]$ is a gaussian white noise independent. It results that
\begin{equation}
 Z_{l,\tau}[n]  \operatornamewithlimits{=}_{Law} \frac{1}{2} \ln \delta_\tau \tilde M_{l,T}(n\tau) + \ln |\epsilon[n]|,
\end{equation} 
and since we will consider below (empirical) covariances of $Z[n]$ for lags $n > 0$, we will not longer take care of the terms $\ln |\epsilon[n] |$.

The empirical covariance function 
involving the cut-off scale $l$ will de denoted as $\widehat{R}_{l,\tau}[n]$:
\begin{equation}
\label{eq:fae1}
\widehat{R}_{l,\tau}[n]=\frac{1}{N}\sum_{k=1}^{N-k}Z_{l,\tau}[k]Z_{l,\tau}[k+n] -\left(\frac{1}{N} \sum_{k=1}^N Z_{l,\tau}[k] \right)^2.
\end{equation}
and naturally
\begin{equation}
\label{eq:fae2}
\widehat{R}_{l,\tau}[n,n']=\widehat{R}_{l,\tau}[n]-\widehat{R}_{l,\tau}[n'].
\end{equation}

A little algebra is sufficient to establish that
\begin{multline}
\label{expvar}
 {\mathbb V}{\mathrm ar}\big[\widehat{R}_{l,\tau}[n,n']\big] = N^{-2} \sum_{j,k} {\mathbb C}{\mathrm ov}\big[Z_{l,\tau}[j]Z_{l,\tau}[j+n],Z_{l,\tau}[k]Z_{l,\tau}[k+n] \big]+ {\mathbb C}{\mathrm ov}\big[Z_{l,\tau}[j]Z_{l,\tau}[j+n'],Z_{l,\tau}[k]Z_{l,\tau}[k+n'] \big] \\ -2 {\mathbb C}{\mathrm ov}\big[Z_{l,\tau}[j]Z_{l,\tau}[j+n],Z_{l,\tau}[k]Z_{l,\tau}[k+n'] \big] 
\end{multline}
Therefore, in order to control the variance of the estimator one needs to control each term
involved in the previous equation. 
Let us introduce some additional notations. Let $[t_0,t_1]$ some time interval such that $t_0/l$ and $t_1/l$ are integers and such that $|t_1-t_0|>2l$. 
For all $t=t_0+jl$ with $0 \leq j \leq (t_1-t_0)/l$,  we will decompose $\omega_{l,T}(t)= {\mathcal P}({\mathcal A}_{l,T}(t))$ as:
\begin{equation}
  \omega_l(t) = O_{[t_0,t_1]} + S_{[t_0,t_1]}^l(j)
\end{equation}
where $O_{[t_0,t_1]}$ and $S_{[t_0,t_1]}^l(j)$ are independant gaussian process defined by
\begin{eqnarray*}
O_{[t_0,t_1]} & = & \frac{1}{2} {\mathcal P}\bigg(\bigcap_{k}{\mathcal A}_{l,T}(t_0+kl)\bigg) \\
S_{[t_0,t_1]}^l(j) & = & \frac{1}{2} {\mathcal P}\bigg({\mathcal A}_{l,T}(t_0+jl) \setminus \bigcap_{k}{\mathcal A}_{l,T}(t_0+kl)\bigg) 
\end{eqnarray*}
Let us now consider two disjoint intervals $[t_0,t_1]$ and $[s_0,s_1$] and let $\delta$ the distance between the middle
of each interval: $\delta = (s_0+s_1-t_0-t_1)/2$. Then after some algebra it can be shown that 
there exists two spherical gaussian vectors\footnote{A spherical gaussian vector is a vector made of independant standard normal variates}
 $d^l_{[t_0,t_1]}$ (of dimension $3(t_1-t_0)/l$) and 
 $r^l_{[s_0,s_1]}$ (of dimension $3(s_1-s_0)/l$) 
such that,
$\forall (j,k) \in [0,(t_1-t_0)/l] \times [0,(s_1-s_0)/l]$,
$S_{[t_0,t_1]}^l(j)$ is a $j$-dependent linear combination of the components of $d^l_{[t_0,t_1]}$ 
while $S_{[s_0,s_1]}^l(k)$ is a $k$-dependent linear 
combination of the components of $r^l_{[s_0,s_1]}$.
Moreover, when $\delta \gg \sup(|s_1-s_0|,|t_1-t_0|)$, one has 
\begin{eqnarray}
\nonumber
\Cov{d^l_{[t_0,t_1]}[k]}{r^l_{[s_0,s_1]}[i]} &= & {\mathcal O}\bigg(\frac{l^2}{\delta^2}\bigg) \; \; \mbox{if} \; \; 0 \leq k \leq (t_1-t_0)/l \; \; \mbox{and}\; \; 0 \leq i \leq (s_1-s_0)/l \\ \nonumber
\Cov{d^l_{[t_0,t_1]}[k]}{r^l_{[s_0,s_1]}[i]} & = & 0 \; \; \mbox{otherwise} \\ \nonumber
\Cov{d^l_{[t_0,t_1]}[k]}{O_{[s_0,s_1]}} &= & {\mathcal O}\bigg(\frac{l}{\delta}\bigg) \; \; \mbox{if} \; \; 0 \leq k \leq (t_1-t_0)/l \\ \label{covnoise}
\Cov{d^l_{[t_0,t_1]}[k]}{O_{[t_0,t_1]}} & = & 0 \; \; \mbox{otherwise} \\ \nonumber
\Cov{r^l_{[t_0,t_1]}[k]}{O_{[t_0,t_1]}} &= & {\mathcal O}\bigg(\frac{l}{\delta}\bigg) \; \; \mbox{if} \; \; 0 \leq i \leq (s_1-s_0)/l \\ \nonumber
\Cov{d^l_{[t_0,t_1]}[k]}{O_{[t_0,t_1]}} & = & 0 \; \; \mbox{otherwise} \\ \nonumber 
\Cov{O_{[t_0,t_1]}}{O_{[s_0,s_1]}} & \sim & R_{\tau}(\frac{\delta}{\tau}) \; \; \nonumber 
\end{eqnarray}   

Let some integer $n > 0$,
according to these notations, $Z_{l,\tau}[k] Z_{l,\tau}[k+n]$ 
can be rewritten as:
\begin{equation}
\label{decomp}
   Z_{l,\tau}[k] Z_{l,\tau}[k+n] = O_{[(k-1)\tau,(k+n)\tau]}^2+ O_{[(k-1)\tau,(k+n)\tau]} f_1(r^l_{[(k-1)\tau,(k+n)\tau]})+f_2(r^l_{[(k-1)\tau,(k+n)\tau]})
\end{equation}
where $f_1$ and $f_2$ are two non-linear functions of the spherical noise $r^l_{[(k-1)\tau,(k+n)\tau]}$ that can be 
shown to have a second moment that is bounded uniformely in $l$.
If one uses the decomposition \eqref{decomp} in expression \eqref{expvar}, each covariance 
term in \eqref{expvar} will give 9 terms that can be of 6 different forms:
If one denotes $I_j$ the interval $[j\tau,(j+n)\tau]$ or  $[j\tau,(j+n')\tau]$:
\begin{eqnarray*}
&& (i) \; N^{-2}{\mathbb C}{\mathrm ov}\big[O_{I_j}^2,O_{I_k}^2] \\
&& (ii) \;  N^{-2}{\mathbb C}{\mathrm ov}\big[O_{I_j}^2,O_{I_k} f_1(r^l_{I_k})\big] \\
&& (iii) \; N^{-2}{\mathbb C}{\mathrm ov}\big[O_{I_j} f_1(d^l_{I_j}),O_{I_k} f_1(r^l_{I_k} \big] \\
 && (iv) \; N^{-2}{\mathbb C}{\mathrm ov}\big[f_2(d^l_{I_j}),f_2(r^l_{I_k}) \big] \\
 && (v) \; N^{-2}{\mathbb C}{\mathrm ov}\big[O_{I_j} f_1(d^l_{I_j}),f_2(r^l_{I_k})  \big] \\
 && (vi) \; N^{-2}{\mathbb C}{\mathrm ov}\big[O_{I_j}^2,f_2(r^l_{I_k}) \big] 
\end{eqnarray*}
In order to prove the consistency of the estimator, we have to prove that  the contribution of each 
term (i)-(vi) vanishes when $N \rightarrow \infty$ (after taking the limit $l \rightarrow 0$).
We will just explain how to take care of the terms of type (iv). 
The other terms are dealt with in the same way. The terms of type (iv) are of the form
\begin{equation}
\label{t6}
   N^{-2} \sum_{j,k} {\mathbb C}{\mathrm ov}\big[O_{I_j}^2,f_2(r^l_{I_k}) \big] 
\end{equation}
We need the following technical Proposition proved in \cite{Robinson} concerning the covariance 
of non-linear function of gaussian vectors:
\begin{Pp}
Let $\mathbf A$ and $\mathbf B$ be two spherical gaussian vectors which dimensions are respectively $p$ and $q$.
Let us denote $C_{kl}$ the cross covariance between components: $C_{kl} = \Cov{A_k}{B_l}$ and
$\rho = \sum_{i=1}^p \sum_{j=1}^q |C_{ij}|$.
Let $f : \RR^p \rightarrow \RR$ and $g : \RR^q \rightarrow \RR$ be two non-linear functions
such that:
\begin{equation}
    \EE{f({\mathbf A})^2} + \EE{g({\mathbf B})^2} < \infty 
\end{equation} 
then, if $\rho < 1/2$ we have:
\begin{equation}
  \Cov{f({\mathbf A})}{g({\mathbf B})} \leq 2 \left(\EE{f({\mathbf A})^2}\EE{g({\mathbf B})^2}\right)^{1/2} \rho
\end{equation}
\end{Pp}
Taking care of the term \eqref{t6} is  a special case of the previous Lemma with $p=1$ and $q=|I_k|/l$. Accordingly and using 
\eqref{covnoise} one has (since $|I_k| \sim \tau$)
\begin{equation}
\rho = \sum_{n=0}^{|I_k|/l} {\mathcal O}(\frac{l}{|j-k| \tau}) = {\mathcal O}(\frac{1}{|j-k|})
\end{equation}
and therefore, there exists $K < \infty$, such that, uniformely in $l$, 
\begin{equation}
{\mathbb C}{\mathrm ov}\big[O_{I_j}^2,f_2(r^l_{I_k}) \big] \leq \frac{K}{|j-k|}
\end{equation}
Let us now remark that in the double sum \eqref{expvar}, since $Z_{\tau} = \ln|\delta_\tau X_T|$ admits a finite four order
moment, one has just to consider the case where $(j,k)$ are such that $|j-k| > N^{\nu}$ with $\nu < 1$ because
the contribution of terms $|j-k| < N^{\nu}$ can be bounded as 
\begin{equation}
   N^{-2} N^{1+\nu} \EE{Z_\tau^4} = {\mathcal O}(N^{\nu-1}) 
\end{equation} 
which converges to zero when $N \rightarrow \infty$. Hence, one can choose some $\nu>0$ in
order that $\sup(n,n') \ll |j-k|$. Under that condition,
the term \eqref{t6} can be bounded, uniformely in $l$ by $ K' \frac{\ln N}{N}$ which
vanishes in the limit $N \rightarrow +\infty$.

The same kind of computation can be lead for each of terms like (ii)-(v) and we do not report
the details here for the sake of concision.
The only problem remains for (i) terms. But in that case, since the random variables $O$ are gaussian wich covariance
is nothing but $R_{\tau}[|j-k|]$, thanks to Wick's theorem the covariance of products of 
$O$ can be expressed in terms of $R_{\tau}$ and it can be shown, that the main remaining contribution
can be bounded like:
\begin{equation}
\frac{2}{N^2}\sum_{|k|<N}(N-|k|)R_{\tau}[k](R_{\tau}[k]-R_{\tau}[k+n-n']
\sim \frac{\ln(N)}{N}.
\end{equation}
whith some constant the depends only on $n$ and $n'$. 
Finally, by merging all contributions together we have proved that 
there exists a constant C that depends only on $n$ and $n'$ such that:
\begin{equation}
 {\mathbb V}{\mathrm ar}\big[\widehat{R}_{l,\tau}[n,n']\big] \leq C \frac{\ln N}{N}
\end{equation}
and by taking the limit $l \rightarrow 0$ one obtains \eqref{eq:majvarMRMbis} and the consistency of the estimator.

\end{IEEEproof}
\newpage

\bibliography{ieee-lncascv4}

\end{document}